\documentclass[%
    prd,      
    preprint, 
    showkeys, 
    superscriptaddress, 
    nofootinbib 
]{revtex4}

\usepackage{amsmath,amsfonts,amssymb}
\usepackage{booktabs} 
\usepackage{csquotes} 
\usepackage{enumitem}
\usepackage{graphicx}
\graphicspath{{./figures/}}

\usepackage{color}
\definecolor{verdes}{cmyk}{0.92,0,0.59,0.4}  
\definecolor{verdec}{cmyk}{0.92,0,0.59,0.15} 

\usepackage{hyperref}
\hypersetup{
    colorlinks=true,
    linkcolor=verdec,
    urlcolor=verdec,
    citecolor=verdec
}

\newcommand{\beq}{\begin{equation}}
\newcommand{\eeq}{\end{equation}}
\newcommand{\bea}{\begin{eqnarray}}
\newcommand{\eea}{\end{eqnarray}}
\newcommand{\barr}{\begin{array}}
\newcommand{\earr}{\end{array}}
\newcommand{\bc}{\begin{center}}
\newcommand{\ec}{\end{center}}
\newcommand{\bit}{\begin{itemize}}
\newcommand{\eit}{\end{itemize}}
\newcommand{\ben}{\begin{enumerate}}
\newcommand{\een}{\end{enumerate}}

\newcommand{\gev}{{\;{\rm GeV}}}

\newcommand{\tauh}{\tau_\text{h}}
\newcommand{\taus}{\tau_\text{sh}}
\newcommand{\kpsw}{\kappa_{\rm\scriptscriptstyle SW}}
\newcommand{\trans}[2]{\text{PT}_{#1}^{(#2\text{-step})}}
\newcommand{\vtr}{V_\text{tree}}
\newcommand{\vcw}{V_\text{CW}}
\newcommand{\vct}{V_\text{CT}}
\newcommand{\vt}{V_T}
\newcommand{\veff}{V_\text{eff}}
\newcommand{\vw}{\vec{w}}
\newcommand{\fz}{\mathcal{F}_0}
\newcommand{\fza}{|\fz|}
\newcommand{\fzsm}{\mathcal{F}_0^\text{SM}}
\newcommand{\fzasm}{|\fzsm|}
\newcommand{\fswp}{f_\text{GW}^\text{SW,peak}}
\newcommand{\omswp}{\Omega_\text{GW}^\text{SW,peak}}
\newcommand{\vwl}{v_\text{w}}
\newcommand{\dfz}{\Delta\mathcal{F}_0}
\newcommand{\aba}{\mathcal{A}_{\alpha\beta}}

\newcommand{\drdmu}{\!\stackrel{\!\!\leftrightarrow}{\partial^\mu}\!}

\newcommand{\br}{\ensuremath\text{Br}}
\newcommand{\hc}{\ensuremath\text{H.c.}}

\newcommand{\fb}{\ensuremath{\,\text{fb}}}
\newcommand{\ab}{\ensuremath{\,\text{ab}}}
\newcommand{\iab}{\ensuremath{\,\text{ab}^{-1}}}
\newcommand{\GeV}{\ensuremath{\,\text{GeV}}}
\newcommand{\TeV}{\ensuremath{\,\text{TeV}}}
\newcommand{\GHz}{\ensuremath{\,\text{GHz}}}
\newcommand{\km}{\ensuremath{\,\text{km}}}

\newcommand{\alg}{\alpha_{\rm\scriptscriptstyle GW}}
\newcommand{\btg}{\beta_{\rm\scriptscriptstyle GW}}
\newcommand{\nsfoewpt}{N_{\rm\scriptscriptstyle SFOEWPT} }
\newcommand{\mh}{m_h}
\newcommand{\mch}{m_{H^\pm}}
\newcommand{\mhh}{m_H}
\newcommand{\ma}{m_A}
\newcommand{\mbsq}{\overline{M}^2}
\newcommand{\mb}{\overline{M}}
\newcommand{\mhsm}{m_{H_\text{SM}}}

\newcommand{\ee}{{e^+ e^-}}
\newcommand{\ttop}{{t\bar{t}}}
\newcommand{\ww}{{W^+ W^-}}

\newcommand{\ca}{c_\alpha}
\newcommand{\sa}{s_\alpha}
\newcommand{\cb}{c_\beta}
\renewcommand{\sb}{s_\beta} 
\newcommand{\tb}{t_\beta}
\newcommand{\cba}{c_{\beta-\alpha}}
\newcommand{\sba}{s_{\beta-\alpha}}

\newcommand{\package}[1]{\textsc{\small #1}}
\newcommand{\filename}[1]{\texttt{\small #1}}

\begin{document}
\preprint{KIAS-P25030}

\title{\color{verdes} Multi-step Strong First-Order Electroweak Phase Transitions in the Inverted Type-I 2HDM: Parameter Space, Gravitational Waves, and Collider Phenomenology} 

\author{Soojin Lee}
\email{soojinlee957@gmail.com}
\affiliation{Department of Physics, Konkuk University, 120 Neungdong-ro, Gwangjin-gu, Seoul 05029, Republic of Korea} 

\author{Dongjoo Kim}
\email{dongjookim.phys@gmail.com}
\affiliation{Department of Physics, Konkuk University, 120 Neungdong-ro, Gwangjin-gu, Seoul 05029, Republic of Korea} 

\author{Jin-Hwan Cho}
\email{chof@nims.re.kr}
\affiliation{National Institute for Mathematical Sciences, Daejeon 34047, Republic of Korea} 

\author{Jinheung Kim}
\email{jhkim1216@kias.re.kr}
\affiliation{School of Physics, Korea Institute for Advanced Study, 85 Hoegi-ro, Dongdaemun-gu, Seoul 02455, Republic of Korea} 

\author{Jeonghyeon Song}
\email{jhsong@konkuk.ac.kr}
\affiliation{Department of Physics, Konkuk University, 120 Neungdong-ro, Gwangjin-gu, Seoul 05029, Republic of Korea} 

\begin{abstract} 
We investigate the electroweak phase transition (EWPT) within the inverted Type-I two-Higgs-doublet model, where the observed $125\GeV$ Higgs boson is identified as the heavier \textit{CP}-even scalar $H$. Through a comprehensive parameter-space scan consistent with current theoretical and experimental constraints, we identify regions supporting strong first-order EWPTs (SFOEWPTs), including multi-step transitions. We find that two-step SFOEWPTs occur as frequently as one-step transitions, while three-step transitions can occur, albeit rarely. Crucially, the parameter spaces inducing one-step and two-step transitions are partially yet significantly separated: one-step transitions restrict the charged Higgs mass and $\tan\beta$ to $m_{H^\pm}\in[295,441]\GeV$ and $\tan\beta\in[4.2,8.8]$, whereas two-step transitions allow $m_{H^\pm}\in[100,350]\GeV$ and $\tan\beta\in[2.5,45.4]$. Notably, negative values of $\sin(\beta-\alpha)$ arise almost exclusively in one-step scenarios. We present the calculation of gravitational wave (GW) signal-to-noise ratios (SNRs) at LISA for multi-step EWPTs, finding that detectable GW signals ($\text{SNR}>10$) predominantly emerge from two-step transitions. Furthermore, we demonstrate that the previously established correlation between the vacuum uplifting measure $\Delta\mathcal{F}_0$ and EWPT strength $\xi_c$ persists only in one-step transitions and breaks down in multi-step cases. Finally, we perform a dedicated collider analysis for representative SFOEWPT parameter points at the $1.5\TeV$ CLIC, identifying $e^+ e^- \to H^+ H^- \to W^+ W^- hh$ as a promising discovery channel. Enhanced $h\to\gamma\gamma$ branching ratios for negative $\sin(\beta-\alpha)$ motivate two complementary golden final states, $W^+ W^- b\bar{b} \tau^+ \tau^-$ and $W^+ W^- b\bar{b}\gamma\gamma$, which demonstrate high discovery potential due to negligible Standard Model backgrounds.
\end{abstract}

\keywords{Electroweak Phase Transition, Higgs Physics, Beyond the Standard Model, Data Analysis}

\maketitle

\tableofcontents

\section{Introduction}

The origin of the observed matter-antimatter asymmetry in the Universe remains one of the most compelling open questions in particle physics.
This baryon asymmetry is quantitatively expressed by the baryon-to-photon ratio, $n_B/n_\gamma \approx 6 \times 10^{-10}$~\cite{Planck:2015fie}, a value far exceeding expectations from a matter-antimatter symmetric early Universe.
The necessary conditions for dynamically generating this asymmetry were outlined by Sakharov~\cite{Sakharov:1967dj}: baryon number violation, \textit{C} and \textit{CP} violation, and a departure from thermal equilibrium.

A strong first-order electroweak phase transition (SFOEWPT) in the early Universe provides a compelling mechanism to satisfy the out-of-equilibrium condition~\cite{Huet:1994jb,Kajantie:1996mn}.
However, for the observed Higgs boson mass of $125\GeV$, the Standard Model (SM) predicts a smooth crossover rather than a first-order transition~\cite{Huet:1994jb,Kajantie:1996mn,Csikor:1998eu}, precluding the possibility of electroweak baryogenesis~\cite{Trodden:1998ym,Cohen:1993nk,Carena:1996wj,Morrissey:2012db} and motivating the search for physics beyond the SM (BSM).

In addition to its crucial role in baryogenesis, an SFOEWPT can also generate a stochastic gravitational wave (GW)  background  through processes such as bubble collisions, plasma sound waves, and turbulence~\cite{Weir:2017wfa,Caprini:2018mtu}.
These GW signals could fall within the sensitivity range of upcoming space-based interferometers like the Laser Interferometer Space Antenna (LISA)~\cite{LISA:2017pwj,Cutting:2018tjt,Guo:2020grp,Schmitz:2020syl}, offering a promising cosmological probe of new physics at the electroweak scale.

Among the well-motivated BSM frameworks capable of inducing an SFOEWPT,\footnote{%
    An SFOEWPT has been investigated in numerous BSM models, including those with an $SU(2)$ singlet~\cite{Carena:2018vpt,Cline:2012hg,Cline:2017qpe,Carena:2018cjh,Cline:2009sn,Profumo:2014opa,Curtin:2014jma,Huang:2015bta,Kotwal:2016tex,Vaskonen:2016yiu,Beniwal:2017eik,Kurup:2017dzf,Chiang:2017nmu,Alves:2018jsw,Li:2019tfd,Bell:2019mbn,Grzadkowski:2018nbc,Huang:2018aja,Ghosh:2022fzp,Roy:2022gop,Azatov:2022tii}, those with an extra triplet~\cite{Inoue:2015pza,Niemi:2018asa,Chala:2018opy,Zhou:2018zli,Kazemi:2021bzj,Addazi:2022fbj,Crivellin:2024uhc,Borah:2024emz,Lu:2025vif,Borah:2025ubr}, 
    composite Higgs models~\cite{Cline:2008hr,Bian:2019kmg,Xie:2020bkl}, 
    supersymmetric models~\cite{Cline:1997vk,Menon:2004wv,Carena:2011jy,Bi:2015qva,Demidov:2016wcv,Huang:2014ifa,Cheung:2012pg,Balazs:2013cia,Huber:2006wf,Bian:2017wfv,Kozaczuk:2014kva,Katz:2015uja,Akula:2017yfr,Lee:2004we,Balazs:2004ae,Liebler:2015ddv,Chatterjee:2022pxf,Borah:2023zsb}, and other frameworks~\cite{Kobakhidze:2015xlz,Ramsey-Musolf:2017tgh,YaserAyazi:2019caf,Mohamadnejad:2019vzg}.
}
the Two-Higgs-Doublet Model (2HDM) remains one of the simplest and most extensively studied~\cite{Bochkarev:1990fx,Dorsch:2013wja,Basler:2016obg,Fuyuto:2017ewj,Bernon:2017jgv,Kainulainen:2019kyp,Bittar:2025lcr}.
The 2HDM extends the SM scalar sector by introducing two $SU(2)_L$ Higgs doublets, $\Phi_1$ and $\Phi_2$~\cite{Branco:2011iw}, leading to a physical spectrum of five Higgs states: two \textit{CP}-even scalars ($h$ and $H$), a \textit{CP}-odd pseudoscalar ($A$), and a pair of charged Higgs bosons ($H^\pm$).
The doublets $\Phi_1$ and $\Phi_2$ acquire vacuum expectation values (VEVs), $v_1$ and $v_2$, with their ratio $\tan\beta = v_2/v_1$ being a key parameter.
To prevent potentially large tree-level Flavor Changing Neutral Currents (FCNCs), a discrete $Z_2$ symmetry, under which $\Phi_1 \to \Phi_1$ and $\Phi_2 \to -\Phi_2$, is typically imposed~\cite{Glashow:1976nt,Paschos:1976ay}.
This leads to four distinct types of Yukawa structures (Type-I, Type-II, Type-X, and Type-Y).

The $125\GeV$ Higgs boson observed at the LHC can be identified with either the lighter scalar $h$ (the normal scenario) or the heavier scalar $H$ (the inverted scenario).
While most studies of SFOEWPTs in the 2HDM have focused on the normal scenario, the inverted scenario is particularly compelling for achieving a strong phase transition.
In the normal scenario, the decoupling limit~\cite{Haber:1989xc,Haber:2006ue,Asner:2013psa,Gunion:2002zf}, characterized by additional Higgs bosons significantly heavier than $125\GeV$, is generally unavoidable, as current and upcoming experiments are unable to probe such heavy scalar masses.
This decoupling limit typically disfavors an SFOEWPT, because the transition strength $\xi_c$ vanishes as the scalar spectrum becomes increasingly decoupled~\cite{Quiros:1999jp,Su:2020pjw,Niemi:2021qvp,Bahl:2024ykv,Masina:2025pnp}, effectively placing an upper bound of around $1\TeV$ on the BSM Higgs masses~\cite{Basler:2016obg,Su:2020pjw}.

In contrast, the inverted scenario inherently avoids a fully decoupled spectrum, since the lighter scalar $h$ necessarily has a mass below $125\GeV$.
However, in inverted scenarios, not all four types remain viable due to stringent experimental constraints. Specifically, Type-II and Type-Y are severely constrained by FCNC processes—most notably from $b \to s\gamma$ transitions—which typically require charged Higgs boson masses $\mch \gtrsim 580\GeV$~\cite{Haller:2018nnx,Misiak:2020vlo,Biekotter:2024ykp,Biekotter:2025fjx}, thus significantly restricting the parameter space compatible with an SFOEWPT.
Consequently, in this work, we focus exclusively on the Type-I 2HDM, demonstrating that this particular inverted scenario can support an SFOEWPT across a substantial region of the parameter space.

An intriguing and relatively unexplored aspect of SFOEWPTs in the inverted 2HDM is the possibility of multi-step phase transitions,\footnote{%
Multi-step EWPTs within the 2HDM framework have primarily been studied in the normal scenario~\cite{Aoki:2021oez,Goncalves:2021egx,Si:2024vrq}.
}
in which the Universe transitions through one or more intermediate metastable vacua before reaching the final true vacuum~\cite{Land:1992sm,Patel:2012pi,Inoue:2015pza,Blinov:2015sna,Bian:2017wfv,Chao:2017vrq,Ramsey-Musolf:2017tgh,Chala:2018opy,Morais:2019fnm,Fabian:2020hny}.
Multi-step transitions are particularly interesting, as they can enhance the strength of individual transition steps~\cite{Angelescu:2018dkk,Morais:2019fnm}, expand the parameter space compatible with an SFOEWPT~\cite{Niemi:2020hto}, and potentially yield multi-peaked GW spectra~\cite{Morais:2019fnm}.
In this work, we will explicitly demonstrate that the inverted Type-I 2HDM accommodates multi-step SFOEWPTs, notably showing that the second transition in a two-step sequence occurs with comparable frequency to conventional one-step transitions.

This naturally raises a crucial question: \textit{are the parameter regions supporting one-step and multi-step SFOEWPTs distinct, or do they overlap?}
If these regions are distinct, measurements of model parameters at high-energy colliders could pinpoint the specific nature of the EWPT, thereby establishing a remarkable interplay between microscopic particle dynamics and macroscopic cosmological events.
To address this question, a comprehensive and detailed scan of the parameter space, fully consistent with current theoretical and experimental constraints, is required.
Through such an extensive parameter-space exploration, we aim to identify viable SFOEWPT parameter regions for each transition type and investigate their combined phenomenological consequences for GW signals and collider observables.

Another important topic closely connected to multi-step EWPTs is the Higgs  vacuum uplifting, characterized by the measure $\dfz$~\cite{Dorsch:2017nza,Su:2020pjw,Goncalves:2021egx}.
Defined explicitly as the difference between $\fz$ in the 2HDM and SM, where $\fz$ represents the potential energy difference between the symmetric and broken phases at zero temperature (see \autoref{subsec-DF0-general} for details), the vacuum uplifting measure has been shown to strongly correlate with the phase transition strength $\xi_c$ in conventional one-step EWPTs.
Consequently, evaluating the zero-temperature potential difference alone is often sufficient to assess the viability of a strong transition in one-step EWPTs.
However, by definition, $\dfz$ does not capture details associated with intermediate metastable vacua present in multi-step EWPTs.
Thus, the previously observed correlation between $\dfz$ and $\xi_c$ might break down  for multi-step transitions, a possibility that has not yet been thoroughly investigated.
Clarifying this issue constitutes another key motivation for the present study.

This paper addresses these open issues by providing a detailed and integrated analysis, with the following novel contributions:
\begin{enumerate}
\item We perform the first systematic mapping of the parameter space compatible with an SFOEWPT in the inverted Type-I 2HDM, including multi-step transitions.
Based on an extensive numerical scan yielding $2.36\times10^6$ physically consistent parameter points, we characterize and distinguish the parameter regions associated with one-, two-, and three-step SFOEWPTs.
\item We investigate the correlation between the vacuum uplifting measure $\dfz$ and the EWPT strength $\xi_c$.
We demonstrate, for the first time, that although this correlation persists for one-step SFOEWPTs, it breaks down in multi-step scenarios.
\item We present the first comprehensive calculation of GW signal-to-noise ratios (SNR) at LISA for multi-step SFOEWPTs in the inverted Type-I 2HDM.
Our analysis reveals that detectable GW signals ($\text{SNR}>10$) predominantly originate from two-step transitions, even though one-step SFOEWPTs are more common in the parameter space.
\item We carry out the first dedicated collider phenomenology analysis of these SFOEWPT-compatible parameter points at the $1.5\TeV$ Compact Linear Collider (CLIC).
We identify the process $e^+e^- \to H^+H^-$, followed by $H^\pm \to W^\pm h$, as a promising discovery channel.
Notably, we find that the decay $h\to\gamma\gamma$ is significantly enhanced for negative $\sba$ values, which correlate strongly with one-step SFOEWPTs.
Based on this result, we propose two complementary golden final states, $W^+W^-b\bar{b}\tau^+\tau^-$ and $W^+W^-b\bar{b}\gamma\gamma$, ensuring high discovery potential for the inverted Type-I 2HDM at the $1.5\TeV$ CLIC.
\end{enumerate}
These interconnected analyses collectively highlight the rich potential of the inverted Type-I 2HDM for generating an SFOEWPT and underscore the critical synergy between early Universe cosmology  and collider experiments in the quest to uncover new physics.

The remainder of this paper is organized as follows. Section~\ref{sec-review} provides a review of the Type-I 2HDM in the inverted Higgs scenario.
In Section~\ref{sec-EWPT}, we detail the formalism for the one-loop effective potential at finite temperature, the dynamics of the EWPT, the Higgs vacuum uplifting measure $\dfz$, and the calculation of GW signals.
Section~\ref{sec-scan} presents our numerical scan methodology, the characteristics of the parameter space for multi-step SFOEWPTs and their GW signals, and the analysis of the  correlation between $\dfz$ and $\xi_c$.
The collider phenomenology at the $1.5\TeV$ CLIC is discussed in Section~\ref{sec-collider}.
Finally, Section~\ref{sec-conclusion} summarizes our main findings and presents our conclusions.
Appendix~\ref{appendix-DF0} contains supplementary details on the calculation of $\dfz$ in the Higgs basis.

\section{Review of Type-I 2HDM in the inverted Higgs scenario}
\label{sec-review}

The 2HDM extends the SM by introducing two complex $SU(2)_L$ scalar doublet fields, $\Phi_1$ and $\Phi_2$, both with hypercharge $Y=1$ under the $Q = T_3 + Y$ convention~\cite{Branco:2011iw}.
These fields are expressed as:
\begin{equation} 
\label{eq:phi:fields}
\Phi_i = \begin{pmatrix} w_i^+ \\ \dfrac{v_i + \rho_i + i \eta_i}{\sqrt{2}} \end{pmatrix}, \quad (i=1,2),
\end{equation}
where $v_1$ and $v_2$ are the VEVs of $\Phi_1$ and $\Phi_2$, respectively.
Their ratio defines a key model parameter, $\tb = v_2/v_1$, where we use $s_x = \sin x$, $c_x = \cos x$, and $t_x = \tan x$ for notational simplicity.
The combined VEV, $v = \sqrt{v_1^2+v_2^2} \approx 246\GeV$, induces spontaneous electroweak symmetry breaking.

To prevent tree-level FCNCs, a discrete $Z_2$ symmetry is imposed, under which $\Phi_1 \to \Phi_1$ and $\Phi_2 \to -\Phi_2$~\cite{Glashow:1976nt,Paschos:1976ay}.
The scalar potential, respecting \textit{CP} invariance and this $Z_2$ symmetry (softly broken), is given by:
\begin{equation}
\label{eq:VPhi}
\begin{split}
V_\Phi (\Phi_1,\Phi_2) &=  m^2_{11} \Phi^\dagger_1 \Phi_1 + m^2_{22} \Phi^\dagger_2 \Phi_2 - m^2_{12} (\Phi^\dagger_1 \Phi_2 + \hc) \\[3pt]
&\quad + \tfrac{1}{2}\lambda_1 (\Phi^\dagger_1 \Phi_1)^2 + \tfrac{1}{2}\lambda_2 (\Phi^\dagger_2 \Phi_2)^2  + \lambda_3 (\Phi^\dagger_1 \Phi_1) (\Phi^\dagger_2 \Phi_2)  \\[3pt]
&\quad + \lambda_4 (\Phi^\dagger_1 \Phi_2)(\Phi^\dagger_2 \Phi_1) + \tfrac{1}{2} \lambda_5 \bigl[ (\Phi^\dagger_1 \Phi_2)^2 + \hc \bigr],
\end{split}
\end{equation}
where the $m_{12}^2$ term softly breaks the $Z_2$ symmetry.
This potential yields five physical Higgs bosons: a lighter \textit{CP}-even scalar $h$, a heavier \textit{CP}-even scalar $H$, a \textit{CP}-odd pseudoscalar $A$, and a pair of charged Higgs bosons $H^\pm$.

The interaction eigenstates are related to the mass eigenstates by rotations involving two mixing angles $\alpha$ and $\beta$~\cite{Song:2019aav}:
\[
\begin{pmatrix} \rho_1 \\ \rho_2 \end{pmatrix} = \mathbb{R}(\alpha) \begin{pmatrix} H \\ h \end{pmatrix},
\quad
\begin{pmatrix} \eta_1 \\ \eta_2 \end{pmatrix} = \mathbb{R}(\beta) \begin{pmatrix} G^0 \\ A \end{pmatrix},
\quad
\begin{pmatrix} w_1^\pm \\ w_2^\pm \end{pmatrix} = \mathbb{R}(\beta) \begin{pmatrix} G^\pm \\ H^\pm \end{pmatrix}.
\]
Here, $G^0$ and $G^\pm$ are the Goldstone bosons absorbed to become the longitudinal components of the $Z$ and $W^\pm$ bosons, respectively.
The rotation matrix $\mathbb{R}(\theta)$ is defined as:
\[
\mathbb{R}(\theta) = \begin{pmatrix} \cos\theta & -\sin\theta \\ \sin\theta & \phantom{-}\cos\theta \end{pmatrix}.
\]

A common and convenient set of independent model parameters, often termed the physical basis, is $
\{ \tb,\cba, \mbsq, v,\mh,\mhh, \ma, \mch\}$ where $\mbsq = m_{12}^2/(\sb\cb)$.
The parameters of $V_\Phi$ in \autoref{eq:VPhi} can be expressed in terms of these physical parameters.
The mass-squared terms $m_{11}^2$ and $m_{22}^2$ are determined by the minimization conditions of the potential:
\[
\begin{split}
  m_{11}^2 &= \sb^2\,\mbsq - \frac{\ca\cba}{2\cb}\mhh^2 + \frac{\sa\sba}{2\cb}\mh^2, \\
  m_{22}^2 &= \cb^2\,\mbsq - \frac{\sa\cba}{2\sb}\mhh^2 - \frac{\ca\sba}{2\sb}\mh^2.
\end{split}
\]
The quartic couplings $\lambda_i$ ($i=1,\dotsc,5$) are then given by:
\[
\begin{split}
  \lambda_1 &= \frac{1}{\cb^2 v^2} \bigl( \sa^2 \mh^2 + \ca^2 \mhh^2 - \sb^2\,\mbsq \bigr), \\
  \lambda_2 &= \frac{1}{\sb^2 v^2} \bigl( \ca^2 \mh^2 + \sa^2 \mhh^2 - \cb^2\,\mbsq \bigr), \\
  \lambda_3 &= \frac{1}{v^2} \bigl[ 2\mch^2 + \frac{s_{2\alpha}}{s_{2\beta}}(\mhh^2-\mh^2) - \mbsq \bigr], \\
  \lambda_4 &= \frac{1}{v^2} \bigl[ \ma^2 - 2\mch^2 + \mbsq \bigr], \\
  \lambda_5 &= \frac{1}{v^2} \bigl[ \mbsq - \ma^2 \bigr].
\end{split}
\]

The SM Higgs boson, $H_\text{SM}$, is a linear combination of $h$ and $H$:
\[
H_\text{SM} = \sba h + \cba H.
\]
In the 2HDM, the Higgs boson observed at the LHC with a mass of $125\GeV$ can be either $h$ (normal scenario) or $H$ (inverted scenario)~\cite{Chang:2015goa,Jueid:2021avn}.
The inverted scenario is of particular interest as theoretical and experimental constraints impose upper bounds on the BSM Higgs boson masses (typically below $\sim 600\GeV$), rendering the model verifiable or falsifiable at high-energy colliders.
This feature makes the inverted scenario particularly compelling for realizing an SFOEWPT, since strong phase transitions are favored in parameter regions with lower scalar masses~\cite{Dorsch:2017nza,Su:2020pjw,Goncalves:2021egx}.
Motivated by these considerations, this work focuses on the inverted Higgs scenario.

The 2HDM has four variants (Type-I, Type-II, Type-X, and Type-Y) based on the $Z_2$ charge assignments to right-handed fermions.
In the inverted Higgs scenario, Type-II and Type-Y are severely constrained and typically do not offer viable parameter space under the theoretical and experimental limits.
This is primarily due to the stringent lower bound on the charged Higgs boson mass ($\mch\gtrsim 580\GeV$) arising from measurements of the inclusive $B$-meson decay $B \to X_s \gamma$~\cite{Haller:2018nnx,Misiak:2020vlo,Biekotter:2024ykp,Biekotter:2025fjx}.
In the context of the inverted scenario (where $m_H = 125\GeV$ and $m_h < 125\GeV$), such a heavy charged Higgs boson leads to large mass splittings within the scalar spectrum.
These large splittings, in turn, severely restrict the allowed parameter space due to theoretical requirements, especially perturbativity and unitarity~\cite{Cheung:2022ndq,Lee:2022gyf}.
Therefore, this paper focuses on the Type-I 2HDM in the inverted Higgs scenario, hereafter referred to as the \enquote{inverted Type-I 2HDM} for simplicity.
Our specific model configuration is thus:
\[
\text{Type-I 2HDM with $\mhh=125\GeV$,}
\]
which features the following six model parameters:
\begin{equation}
\label{eq-model-parameters}
\big\{ \tb,\; \cba,\; \mbsq,\;  \mh,\;  \ma,\; \mch \big\}.
\end{equation}

The interactions between the Higgs bosons and gauge bosons are described by the Lagrangian term:
\begin{equation}
\label{eq-gauge-coupling}
\begin{split}
\mathcal{L}_\text{gauge} &=
\Bigl( g m_W W^\dagger_\mu W^\mu + \tfrac{1}{2} g_Z m_Z Z_\mu Z^\mu \Bigr)
\Bigl( \sba h + \cba H \Bigr) \\
&\quad + \dfrac{g}{2} i
\Bigl[ W_\mu^+ ( \cba h -\sba H ) \drdmu H^- - \hc \Bigr]
- \dfrac{g}{2} \Bigl[ W_\mu^+  H^- \drdmu A  + \hc \Bigr] \\
&\quad + i
\Bigl\{ e A_\mu + \dfrac{g_Z}{2}(s_W^2-c_W^2) Z_\mu \Bigr\} H^+ \drdmu H^-
+ \dfrac{g_Z}{2} Z_\mu \Bigl[ \cba A \drdmu h -\sba A \drdmu H \Bigr],
\end{split}
\end{equation}
where $s_W = \sin\theta_W$ (with $\theta_W$ denoting the electroweak mixing angle), $g_Z = g/c_W$, and the shorthand $f \drdmu g$ is defined as $f \drdmu g \equiv f(\partial^\mu g) - (\partial^\mu f)g$.

The Yukawa Lagrangian is parametrized as:
\[
\begin{split}
\mathcal{L}_\text{Yuk} &=  - \sum_f \Bigl(
\frac{m_f}{v} \xi^h_f \bar{f} f h + \frac{m_f}{v} \kappa^H_f \bar{f} f H - i \frac{m_f}{v} \xi^A_f \bar{f} \gamma_5 f A 
\Bigr) \\
&\quad - \Bigl\{
\dfrac{\sqrt{2}V_{ud}}{v} H^+ \overline{u} (m_u \xi^A_u \text{P}_L +  m_d \xi^A_d \text{P}_R)d
+ \dfrac{\sqrt{2} m_\ell}{v} H^+ \xi^A_\ell \overline{\nu}_L\ell_R  + \hc
\Bigr\}.
\end{split}
\]
In the Type-I 2HDM, the Yukawa coupling modifiers are:
\begin{equation}
\label{eq-Yukawa}
\kappa^H_f = \frac{\sa}{\sb} = \cba - \frac{\sba}{\tb}, \quad
\xi^h_f = \frac{\ca}{\sb} = \sba + \frac{\cba}{\tb}, \quad
\xi^A_t = -\xi^A_{b,\tau} = \frac{1}{\tb}.
\end{equation}

\section{One-Loop Effective Potential, Electroweak Phase Transition Dynamics, and GW signals}
\label{sec-EWPT}

\subsection{One-Loop Effective Potential}

The dynamics of the EWPT in the early universe is governed by the effective potential at finite temperature~\cite{Espinosa:1992kf,Herring:2024pqa,Chakrabortty:2024wto,Masina:2025pnp}.
This potential is typically evaluated for spatially homogeneous field configurations.\footnote{%
    For studying the EWPT as a cosmological phase transition driven by scalar field dynamics in thermal equilibrium, it is appropriate to consider the effective potential as a function of spatially homogeneous background fields.
    This approach is consistent with the homogeneous and isotropic nature of the cosmological background.
}
We therefore replace the quantum fields $\Phi_{1,2}$ (defined in \autoref{eq:phi:fields}) with their classical, constant background configurations $\Phi^c_{1,2}$, expressed in terms of three homogeneous real scalar fields $w_1$, $w_2$, and $w_3$ as:
\[
\Phi_1^c = \frac{1}{\sqrt{2}} \begin{pmatrix} 0 \\ w_1 \end{pmatrix}, \quad
\Phi_2^c = \frac{1}{\sqrt{2}} \begin{pmatrix} 0 \\ w_2 + i w_3 \end{pmatrix}.
\]
The inclusion of $w_3$, even in a \textit{CP}-conserving 2HDM where the zero-temperature vacuum preserves \textit{CP} symmetry, is essential to account for the possibility of a \textit{CP}-violating vacuum state or path at finite temperature~\cite{Ferreira:2015pfi,Basler:2016obg,Si:2024vrq}.
The full one-loop effective potential at a temperature $T$ is then given by the sum of four components:
\beq
\label{eq-Veff}
\veff(\vw, T) = \vtr(\vw) +\vcw(\vw) + \vct(\vw) + \vt(\vw, T),
\eeq
where $\vw=(w_1,w_2,w_3)$.
The tree-level potential $\vtr(\vw)$ is obtained by substituting $\Phi_{1,2} \rightarrow \Phi^c_{1,2}$ into the 2HDM potential $V_\Phi(\Phi_1, \Phi_2)$ given in \autoref{eq:VPhi}.

The one-loop Coleman-Weinberg (CW) potential $\vcw(\vw)$ in the $\overline{\text{MS}}$ renormalization scheme is given by~\cite{Coleman:1973jx}:
\begin{equation}
\label{eq-VCW}
\vcw(\vw) = \frac{1}{64 \pi^2} \sum_i n_i\, m_i^4(\vw) \Bigl[ \ln\Bigl( \frac{m_i^2(\vw)}{\mu^2} \Bigr) - c_i \Bigr],
\end{equation}
where the sum extends over all massive particle species $i = h, H, A, H^\pm, G^0, G^\pm, W^\pm, Z, f$, with $f$ representing the SM fermions.
The factor $n_i$ denotes the number of degrees of freedom for each species, being positive for bosons and negative for fermions.
Specifically, $n_i = -12$ for each quark, $-4$ for each charged lepton, $6$ for $W^\pm$ bosons, $3$ for the $Z$ boson, $2$ for $H^\pm$, $2$ for $G^\pm$, $1$ for $G^0$, and $1$ for each real neutral scalar ($h, H, A$).

The field-dependent squared masses, $m_i^2(\vw)$, are determined by diagonalizing the field-dependent mass matrices for all particles that couple to the scalar background fields $\vw$.
Their explicit forms for the 2HDM can be found in, e.g., Ref.~\cite{Basler:2016obg}.
The renormalization scale $\mu$ is set to the electroweak scale $v \approx 246\GeV$.
The constants $c_i$'s depend on the particle type in the $\overline{\text{MS}}$ scheme: $c_i = 5/6$ for gauge bosons and $c_i = 3/2$ for scalars and fermions.

Although the CW potential in \autoref{eq-VCW} incorporates $\overline{\text{MS}}$ subtraction, an on-shell renormalization scheme is adopted by introducing a counter-term potential, $\vct(\vw)$.
This ensures that the physical input parameters in \autoref{eq-model-parameters} correspond to their intended physical values at the one-loop level, thus preventing shifts in observables.
The counter-term potential has the  form of
\begin{equation}
\label{eq:VCT}
\begin{split}
\vct(\vw) &= 
    \delta m_{11}^2 \frac{w_1^2}{2} +
    \delta m_{22}^2 \frac{w_2^2 + w_3^2}{2} -
    \delta m_{12}^2 w_1 w_2 +
    \frac{\delta\lambda_1}{8} w_1^4 + 
    \frac{\delta\lambda_2}{8} (w_2^2 + w_3^2)^2 \\
&\quad + \frac{\delta\lambda_3 +
    \delta\lambda_4}{4} w_1^2 (w_2^2 + w_3^2) +
    \frac{\delta\lambda_5}{4} w_1^2(w_2^2 - w_3^2).
\end{split}
\end{equation}
Explicit expressions for these counter-terms in the Landau gauge are provided in Ref.~\cite{Basler:2016obg}, where techniques are also discussed for regulating the divergent logarithmic contributions from massless Goldstone bosons, which are characteristic of this gauge choice.

Finally, the thermal correction $\vt(\vw, T)$~\cite{Dolan:1973qd,Quiros:1999jp}, including daisy resummation for thermal masses~\cite{Carrington:1991hz}, is given by:
\begin{equation}
\label{eq-VT}
\begin{split}
\vt(\vw, T) &= 
	\frac{T^4}{2\pi^2} \sum_{i_B} n_{i_B} J_B \Bigl( \frac{m^2_{i_B}(\vw)}{T^2} \Bigr) +
	\frac{T^4}{2\pi^2} \sum_{j_F} n_{j_F} J_F \Bigl( \frac{m^2_{j_F}(\vw)}{T^2} \Bigr) \\
&\quad - \frac{T^4}{12\pi} \sum_{k_L} n_{k_L} \biggl[
		\Bigl( \frac{\tilde{m}_{k_L}^2(\vw, T) }{T^2} \Bigr)^{3/2} -
		\Bigl( \frac{m_{k_L}^2(\vw)}{T^2} \Bigr)^{3/2}
	\biggr],
\end{split}
\end{equation}
where the sum over $i_B$ includes all bosonic species (scalars and gauge bosons), $j_F$ includes all fermionic species, and the sum over ${k_L}$ in the daisy resummation term includes scalars and the longitudinal components of gauge bosons that acquire thermal masses.
The thermal functions $J_B(x)$ and $J_F(x)$ are defined by the standard integrals:
\[
J_B(x) = \int_0^\infty dk \, k^2 \ln \bigl[ 1 - e^{-\sqrt{k^2 + x}} \bigr], \quad
J_F(x) = \int_0^\infty dk \, k^2 \ln \bigl[ 1 + e^{-\sqrt{k^2 + x}} \bigr].
\]
The temperature-dependent thermal (Debye) masses, $\tilde{m}_k^2(\vw, T)$, appearing in the daisy resummation term in \autoref{eq-VT}, are crucial for improving the perturbative expansion at high temperatures and regulating infrared divergences associated with light scalar modes.
Explicit expressions for $\tilde{m}_k^2(\vw, T)$ can be found in the literature such as Refs.~\cite{Arnold:1992rz,Basler:2016obg}.

\subsection{Thermal History and Transition Dynamics}
\label{sec:ThermalHistory}

The dynamics of the EWPT in the early Universe is governed by the loop-corrected effective potential, $\veff(w_1, w_2, w_3, T)$.
At very high temperatures, long before electroweak symmetry breaking, this potential features a single global minimum at $w_1 = w_2 = w_3 = 0$, corresponding to the electroweak symmetric vacuum.
As the Universe cools, additional local minima may develop at nonzero field values.
The coordinates of such a minimum evolve as a function of temperature and are denoted by $(v_1^T, v_2^T, v_3^T)$.

At a particular temperature $T_c$, one of these broken-phase minima may become degenerate in free energy with the symmetric minimum.
This temperature is known as the critical temperature and is defined by the condition:
\begin{equation}
\label{eq-Tc-1step}
\veff(0,0,0, T_c) = \veff(v_1^{T_c}, v_2^{T_c}, v_3^{T_c}, T_c).
\end{equation}

As the temperature continues to decrease below $T_c$, this broken-phase minimum  becomes energetically favored over the symmetric vacuum.
If a potential barrier separates this now true (broken) vacuum from the false (symmetric) vacuum, the transition proceeds via thermal tunneling—a hallmark of a \emph{first-order phase transition} (FOEWPT).
A key measure of such a transition's strength is the order parameter $\xi_c$, defined at the critical temperature $T_c$ as:
\begin{equation}
\label{eq-xic-one-step}
\xi_c \equiv \frac{v_c}{T_c},
\end{equation}
where $v_c = \sqrt{\sum_{i=1}^3 (v_i^{T_c})^2}$ is the magnitude of the VEV in the broken phase at $T_c$.
These definitions of $T_c$ and $\xi_c$ apply directly to a one-step transition originating from the symmetric phase.
However, the EWPT may proceed through multiple steps, involving a sequence of transitions via intermediate metastable vacua before the Universe reaches its final true vacuum~\cite{Land:1992sm,Patel:2012pi,Inoue:2015pza,Blinov:2015sna,Bian:2017wfv,Chao:2017vrq,Ramsey-Musolf:2017tgh,Chala:2018opy,Morais:2019fnm,Fabian:2020hny}.
In such multi-step scenarios, the definitions of the critical temperature and order parameter for each individual step are modified, as will be detailed in \autoref{subsec-scan-results}.

For successful electroweak baryogenesis, the FOEWPT typically needs to be strong in order to sufficiently suppress sphaleron-induced baryon number washout in the broken phase. An SFOEWPT is often quantified by the condition $\xi_c > 1$,  
In this work, we adopt a more stringent criterion~\cite{Quiros:1999jp}:
\bea
\label{eq-xic-1.3}
\xi_c > 1.3.
\eea
This choice is motivated by our analysis, which shows that achieving an SNR greater than 10 for detectable GWs at LISA frequently requires large values of $\xi_c$, often exceeding 2.  
Thus, imposing $\xi_c > 1.3$ helps isolate regions of the 2HDM parameter space that are more likely to yield both an SFOEWPT and observable GW signatures.

The actual realization of a FOEWPT, triggered by thermal tunneling, hinges on the nucleation of true vacuum bubbles that are large enough to expand rather than collapse.  
This occurs at the nucleation temperature $T_n$, where critical bubbles---those that have just overcome the energy barrier associated with surface tension---become energetically favored to grow.

The nucleation temperature $T_n$ is formally defined by the condition that  the probability to nucleate a bubble within one Hubble volume during one Hubble time, $\Gamma/H^4$, equals unity~\cite{Coleman:1977py,Goncalves:2021egx}:
\[
\frac{\Gamma(T_n)}{H^4(T_n)} = 1,
\]
where $\Gamma(T)$ is approximately given by
\[
\Gamma(T) \simeq T^4 \Bigl( \frac{S_3(T)}{2\pi T} \Bigr)^{3/2} \exp\Bigl( -\frac{S_3(T)}{T} \Bigr).
\]
Here, $S_3(T)$ is the three-dimensional Euclidean action for the $O(3)$-symmetric bounce solution connecting the false and true vacua:
\[
S_3(T) = 4\pi \int_0^\infty dr\, r^2 \biggl[ \frac{1}{2} \sum_k \Bigl( \frac{d w_k(r)}{dr} \Bigr)^2 + \veff(\vw(r), T) \biggr],
\]
where $\vw(r)$ with components $w_k(r)$ represents the field profiles along the radial coordinate $r$ for the fields participating in the bounce.
These profiles are solutions to the bounce equations:
\[
\frac{d^2 w_k}{dr^2} + \frac{2}{r} \frac{dw_k}{dr} = \frac{\partial \veff(\vw(r),T)}{\partial w_k}, 
\]
subject to the boundary conditions:
\[
\lim_{r \to \infty} \vw(r) = \vw_{\text{false}}, \quad \frac{d\vw(r)}{dr}\Big|_{r=0} = \vec{0},
\]
where $\vw_\text{false}$ denotes the field configuration in the false vacuum.
In practice, $T_n$ is often well-approximated by the criterion $S_3(T_n)/T_n \simeq 140$~\cite{Linde:1980tt,Coleman:1977py}.

Following their nucleation, bubbles of the true vacuum expand.
As these bubbles grow and coalesce, they eventually percolate, filling space and thereby completing the FOEWPT.
The percolation temperature $T_p$ is defined as the temperature at which at least 29\% of the false vacuum has tunneled into the true vacuum.
This temperature marks the practical completion of the phase transition, with the Universe having largely transitioned out of the symmetric phase into the lower-energy state.

Finally, at zero temperature, the present-day Universe resides in the global minimum of the effective potential.
In the \textit{CP}-conserving 2HDM, this minimum corresponds to the VEVs:
\[ 
v_1^{T=0} = v \cb, \quad v_2^{T=0}  = v \sb, \quad v_3^{T=0}  = 0.
\]

\subsection{Higgs Vacuum Uplifting and Its Implications}
\label{subsec-DF0-general}

An important theoretical diagnostic for the nature of the EWPT is the Higgs vacuum uplifting~\cite{Huang:2014ifa,Harman:2015gif,Dorsch:2017nza}.
This phenomenon is characterized by the difference in zero-temperature vacuum energy density between the 2HDM and the SM.
To quantify this, we first express the one-loop effective potential at temperature $T$ as:
\[
\veff(\vw, T) = V_0(\vw) + \vt(\vw, T),
\]
where the zero-temperature component $V_0(\vw)$ is
\[ 
V_0(\vw) = \vtr(\vw) + \vcw(\vw) + \vct(\vw).
\]

The vacuum energy density difference, $\mathcal{F}_T$, between the symmetric phase where $\langle |\vw| \rangle = 0$ and a broken phase where $\langle |\vw| \rangle = v_T$ at temperature $T$ is given by:
\bea
\label{eq-FT}
\begin{split}
\mathcal{F}_T &= \veff(v_T, T) - \veff(0, T) \\ 
&= \bigl[ V_0(v) - V_0(0) \bigr] + \bigl[ V_0(v_T) - V_0(v) + \vt(v_T, T) - \vt(0, T) \bigr] \\
&\equiv \fz + \Delta V_T,
\end{split}
\eea
where $\fz \equiv V_0(v) - V_0(0)$ is the vacuum energy density difference between the symmetric and broken phases at zero temperature.
Since a successful EWPT requires $\fz < 0$, we define $\fz = -\fza$.
The thermal correction term, $\Delta V_T$, typically increases monotonically with temperature.

A key quantity in analyzing vacuum uplifting is the difference  between $\fz$ in the 2HDM and $\fz$ in the SM:
\bea
\label{eq-DtF0}
\dfz \equiv \fz^\text{2HDM} - \fzsm,
\eea
where $\fzsm \approx -1.25 \times 10^8\GeV^4$.
$\dfz$ is particularly useful as it is renormalization-scale independent and  gauge invariant, providing a robust theoretical measure.

In the 2HDM, $\dfz$ is given by~\cite{Dorsch:2017nza}:
\beq
\label{eq-dfz}
\begin{split}
\dfz &= \dfz^\text{tree} - \frac{m_{H_\text{SM}}^4}{64\pi^2}(3 + \log 2) 
    -\sum_{k} \frac{m_{0_k}^4}{64\pi^2} \Bigl( \log\frac{|m_{0_k}^2|}{\mu^2} - \frac{1}{2} \Bigr) \\
&\quad + \frac{1}{64\pi^2} \sum_k \frac{1}{4} \biggl\{
	\tilde{I}_k^2 - 2\,m_k^4 + \Bigl[
    	\bigl( \tilde{I}_k - 2\,m_k^2 \bigr)^2 + m_k^2\,\bigl( \tilde{J}_k - \tilde{I}_k \bigr)
	\Bigr] \log\frac{m_k^2}{\mu^2} \biggr\},
\end{split}
\eeq
where the sum over $k$ includes the scalar states of $H^\pm, A, H, h, G^\pm, G^0$.
The expressions for $m_{0_k}^2$ (the scalar squared masses at the origin, i.e., in the symmetric vacuum) and the field-derivative quantities $\tilde{I}_k$ and $\tilde{J}_k$ are detailed in Appendix~\ref{appendix-DF0}.
The tree-level contribution $\dfz^\text{tree}$ depends on whether the normal or inverted Higgs scenario is considered:
\begin{equation}
\label{eq-dfz-tree}
\dfz^{\text{tree}} = \begin{cases}
	- \cba^2\dfrac{v^2(\mhh^2-\mh^2)}{8}, & \text{if $\mh=125\GeV$ (normal scenario);} \\[9pt]
	+ \sba^2\dfrac{v^2(\mhh^2-\mh^2)}{8}, & \text{if $\mhh=125\GeV$ (inverted scenario).}
\end{cases}
\end{equation}
In the normal Higgs scenario, $\dfz^\text{tree}$ is negative definite, while in the inverted Higgs scenario, it is positive definite.
In both scenarios, $\dfz^\text{tree}$ vanishes in the Higgs alignment limit where either $\sba=0$ (Normal) or $\cba=0$ (Inverted).

In the Higgs alignment limit of the inverted 2HDM, corresponding to $\sba=0$, $\dfz$ simplifies to
\begin{equation}
\label{eq-dfz-sba0}
\begin{split}
\dfz\Big|_{\sba=0} &= \frac{1}{64\pi^2} \biggl[
	(\mhh^2-2\mbsq)^2 \, \biggl\{ \frac{3}{2} + \frac{1}{2} \log \frac{4\ma\mh\mch^2}{\bigl(\mhh^2-2\mbsq\bigr)^2} \biggr\} \\
&\qquad + \frac{1}{2} (\ma^4 + \mh^4 + 2\mch^4) + (\mhh^2 - 2\mbsq)(\ma^2 + \mh^2 + 2\mch^2) \biggr],
\end{split}
\end{equation}
which is generally positive definite.

It was first demonstrated through empirical studies in Ref.~\cite{Dorsch:2017nza} that, in the 2HDM, the quantity $\dfz$ exhibits a strong correlation with the strength of the EWPT, as measured by $\xi_c$.  
This correlation makes the analysis of $\dfz$ an efficient approach since calculations of this zero-temperature quantity is analytically simpler than calculations involving the full thermal effective potential.
Specifically, for one-step transitions in the normal Higgs scenario, it was found that an SFOEWPT is typically guaranteed if
\begin{equation}
\label{eq-dfz-onestep}
\text{SFOEWPT via one-step with $\mh = 125\gev$:} \quad 0.34 \lesssim \frac{\dfz}{\fzasm} < 1.
\end{equation}
The upper bound ensures that the 2HDM electroweak vacuum remains the global minimum at $T=0$.
This range can be reinterpreted  as $0 < |\fz^\text{2HDM}| \lesssim 0.66 \fzasm$.
This condition implies that for an SFOEWPT, the depth of the true vacuum at $T=0$ (relative to the symmetric phase) becomes shallower compared to the depth in the SM, meaning that the Higgs vacuum is effectively \enquote{uplifted}.
Consequently, a more significant uplifting of the Higgs potential at zero temperature tends to correlate with a stronger FOEWPT.
In light of this, we will refer to $\dfz$ as the Higgs vacuum uplifting measure in subsequent discussions.
 
\subsection{Gravitational Wave Signal}
\label{subsec-GW}

An SFOEWPT involves the nucleation and expansion of true vacuum bubbles, which subsequently grow, collide, and coalesce, ultimately completing the transition throughout the Universe.  
These highly energetic processes can leave cosmological imprints, making SFOEWPTs an intriguing target for probing early Universe dynamics~\cite{Weir:2017wfa,Tenkanen:2022tly}.  
If the transition is sufficiently strong, its signatures may be detectable at future space-based interferometers such as LISA, offering a unique cosmological probe of the early Universe~\cite{Caprini:2019egz,Caldwell:2022qsj,Auclair:2022lcg}.

The GW spectrum produced in a FOEWPT is characterized by four essential quantities~\cite{Caprini:2019egz,Auclair:2022lcg,Basler:2024aaf}:
    (i) the transition temperature $T_*$;
    (ii) the strength of the phase transition $\alg$; 
    (iii) the inverse duration of the phase transition in Hubble units at the transition temperature, $\btg/H_*$; and
    (iv) the bubble wall velocity $\vwl$.
Although the notations $\alpha$ and $\beta$ are commonly used in the literature for the GW spectrum parameters, we adopt $\alg$ and $\btg$ herein to avoid confusion with the mixing angles $\alpha$ and $\beta$ of the 2HDM.

The first important quantity is the transition temperature $T_*$, which is the characteristic temperature at which thermal parameters are evaluated for predicting GW signals from a FOEWPT.
In this work, we choose $T_*$ to be the percolation temperature $T_p$.

The second parameter $\alg$ quantifies the strength of the phase transition and is defined by~\cite{Hindmarsh:2015qta,Hindmarsh:2017gnf}:
\[
\alg = \frac{\epsilon}{\rho_\gamma},
\]
where $\epsilon$ is the latent heat released during the phase transition, and $\rho_\gamma = g_* \pi^2 T_*^4/30$ is the radiation energy density at $T_*$.
Here, $g_* \equiv g_*(T_*)$ is the effective number of relativistic degrees of freedom in the plasma (for the 2HDM context of this study).
As a reference, in the SM, $g_* \simeq 106.75$~\cite{Caprini:2015zlo,Caprini:2019egz,Grojean:2006bp,Leitao:2015fmj}.

The third key parameter $\btg/H_*$ represents the inverse duration of the phase transition relative to the Hubble time at $T_*$, and is defined as~\cite{Ellis:2018mja}:
\[
\frac{\btg}{H_*} \equiv \left. T_* \frac{d}{dT} \left(\frac{S_3}{T}\right)\right|_{T=T_*},
\]
where $H_*$ is the Hubble parameter at temperature $T_*$:
\[ 
H_* = T_*^2 \sqrt{ \frac{g_*\pi^2}{90 M_{\rm Pl}^2} },
\]
with the reduced Planck mass $M_{\rm Pl} \approx 2.4\times 10^{18}\GeV$~\cite{Ellis:2018mja}.

The final parameter $\vwl$ is the speed of the bubble wall after nucleation, measured in the rest frame of the plasma.
This parameter critically influences the GW spectrum by affecting the efficiency of energy transfer from the phase transition to the surrounding plasma, thereby shaping both the peak amplitude and peak frequency of the resulting GW signal~\cite{Ai:2021kak,Dorsch:2021nje,Jiang:2022btc,DeCurtis:2022hlx,Ai:2023see,DeCurtis:2023hil,Krajewski:2024gma,Wang:2024wcs,DeCurtis:2024hvh,Branchina:2025jou}.
Despite promising recent proposals~\cite{Lewicki:2021pgr,Laurent:2022jrs}, a universally accepted prediction for $\vwl$ from first principles is still lacking, and its determination remains highly intricate and model-dependent.
Consequently, in our analysis, we assume a constant terminal bubble wall velocity---a reasonable approximation for non-runaway transitions where the acceleration phase is brief compared to the bubble lifetime---and thus treat $\vwl$ as an input parameter.

In the 2HDM, FOEWPTs with non-runaway bubble expansion produce GWs predominantly from sound waves~\cite{Biekotter:2022kgf,Ramsey-Musolf:2024zex}.\footnote{We scanned the parameter space for SFOEWPT-viable points and computed the corresponding GW spectra.  
Our results, presented in the next section, show that turbulence contributes negligibly compared to sound waves in the inverted Type-I 2HDM.}
The GW power spectrum from the dominant sound wave contribution is given by~\cite{Grojean:2006bp,Leitao:2015fmj,Caprini:2001nb,Figueroa:2012kw,Caprini:2015zlo,Hindmarsh:2016lnk}:
\begin{equation}
\label{eq-GW-spectrum}
h^2\,\Omega_\text{GW}(f) \simeq h^2 \, \omswp \Bigl(\frac{4}{7}\Bigr)^{-\frac{7}{2}} \Bigl(\frac{f}{\fswp}\Bigr)^3
	\biggl[ 1 + \frac{3}{4} \Bigl(\frac{f}{\fswp}\Bigr)^2 \biggr]^{-\frac{7}{2}},   
\end{equation}
where $h=0.674\pm 0.005$~\cite{ParticleDataGroup:2024cfk}, $\omswp$ and $\fswp$ are the peak amplitude and peak frequency of the sound wave contribution, respectively.

The peak frequency $\fswp$ is given by:
\[
\fswp = 8.88 \times 10^{-6}\,\text{Hz}\,
    \Bigl(\frac{g_*}{100}\Bigr)^{\frac{1}{6}}
    \Bigl(\frac{T_*}{100\GeV}\Bigr) 
    \Bigl(\frac{\btg}{H_*}\Bigr)
	\frac{1}{\max(\vwl, c_s)},
\]
where $c_s = 1/\sqrt{3}$ is the speed of sound in the relativistic plasma.
Semi-analytic expressions for the peak amplitude $h^2 \omswp$ are provided in Refs.~\cite{Caprini:2019egz,Hindmarsh:2017gnf,Athron:2023xlk} as:
\begin{align}
\label{eq-omega-peak}
&  h^2 \omswp \\ \nonumber
&= \begin{cases} 
8.75 \times 10^{-6} h^2  
	\Bigl(\dfrac{100}{g_*}\Bigr)^{\frac{1}{3}} 
	\Bigl(\dfrac{H_*}{\btg}\Bigr)^2  
	\Bigl(\dfrac{\kpsw\alg}{1+\alg}\Bigr)^{\frac{3}{2}}  
	\max(\vwl, c_s)^2, & \text{if $H_*\taus<1$;} \\[10pt]
2.59 \times 10^{-6} h^2
	\Bigl(\dfrac{100}{g_*}\Bigr)^{\frac{1}{3}} 
	\Bigl(\dfrac{H_*}{\btg}\Bigr)
	\Bigl(\dfrac{\kpsw\alg}{1+\alg}\Bigr)^2  
	\max(\vwl, c_s), & \text{if $H_*\taus\geq 1$,}
\end{cases}
\end{align}
where  the maximum effective lifetime of the sound wave source is taken to be one Hubble time, $\tau_{\rm sw} = H_*^{-1}$~\cite{Athron:2020sbe}.

In \autoref{eq-omega-peak}, $\taus$ is the shock formation timescale, and $H_* \taus$ is given by:
\[
H_* \taus = \frac{4 \pi^{\frac{1}{3}}}{\sqrt{3}}
	\Bigl(\dfrac{H_*}{\btg} \Bigr)
	\Bigl(\dfrac{\kpsw\alg}{1+\alg}\Bigr)^{-\frac{1}{2}}
	\max(\vwl, c_s).
\]
The quantity $\kpsw$ is the sound wave efficiency factor, given by~\cite{Espinosa:2010hh}:
\beq
\label{eq-kpsw}
\kpsw = \begin{cases}
	\dfrac{c_s^{11/5} \kappa_A \kappa_B}{\bigl(c_s^{11/5}-\vwl^{11/5}\bigr) \kappa_B + \vwl c_s^{6/5} \kappa_A}, & \text{if $\vwl<c_s$;} \\[15pt]
	\kappa_B + (\vwl-c_s) \delta \kappa + \dfrac{(\vwl-c_s)^3}{(v_J-c_s)^3} \bigl[\kappa_C-\kappa_B-(v_J-c_s)\delta\kappa\bigr], & \text{if $c_s<\vwl<v_J$;} \\[15pt]
	\dfrac{(v_J-1)^3 v_J^{5/2} \vwl^{-5/2} \kappa_C \kappa_D}{\bigl[(v_J-1)^3-(\vwl-1)^3\bigr] v_J^{5/2} \kappa_C+(\vwl-1)^3 \kappa_D}, & \text{if $v_J<\vwl$,} 
\end{cases}
\eeq
where  $v_J$ is the Chapman-Jouguet velocity~\cite{Steinhardt:1981ct,Kamionkowski:1993fg,Espinosa:2010hh}, defined as:
\[
v_{J} = \frac{1}{1+\alg} \Bigl( c_s + \sqrt{\alg^2 + \tfrac{2}{3} \alg} \Bigr).
\]
The auxiliary functions $\kappa_X$ (for $X=A,B,C,D$) and $\delta\kappa$ in \autoref{eq-kpsw} are:
\[
\begin{alignedat}{2}
    \kappa_A &\simeq \vwl^{6 / 5} \frac{6.9 \,\alg}{1.36-0.037\sqrt{\alg}+\alg}, &\quad  
    	\kappa_B &\simeq \frac{\alg^{2 / 5}}{0.017+\left(0.997+\alg\right)^{2 / 5}}, \\
    \kappa_C &\simeq \frac{\sqrt{\alg}}{0.135+\sqrt{0.98+\alg}}, &\quad
    	\kappa_D &\simeq \frac{\alg}{0.73+0.083 \sqrt{\alg}+\alg}, \\
    \delta \kappa &\simeq -0.9 \log \frac{\sqrt{\alg}}{1+\sqrt{\alg}}. &\quad
\end{alignedat}
\]

The detectability of GW signals at the future LISA is quantified by the SNR, given by~\cite{Caprini:2019egz}:
\[
\text{SNR} = \sqrt{\mathcal{T} \int_{f_\text{min}}^{f_\text{max}} df \Bigl[\frac{h^{2}\,\Omega_\text{GW}(f)}{h^{2}\,\Omega_\text{Sens}(f)}\Bigr]^2}.
\]
Here, $\mathcal{T}$ is the experimental observation time in seconds, taken as four years for LISA in this study~\cite{LISA:2017pwj}.
The integration limits $f_\text{min}$ and $f_\text{max}$ define LISA's sensitive frequency range.
The term $h^{2}\,\Omega_\text{GW}(f)$ is the predicted GW signal (from the 2HDM in this work), and $h^2\,\Omega_\text{Sens}(f)$ is the nominal sensitivity of the LISA configuration to stochastic backgrounds~\cite{Caprini:2019pxz,Babak:2021mhe}.
A signal is typically considered detectable if $\mathrm{SNR} \gtrsim 10$~\cite{Caprini:2015zlo}.

\section{Multi-step EWPT: Parameter Space, Vacuum Structure, and Gravitational Waves}
\label{sec-scan}

\subsection{Scan-Based Analysis of Multi-step SFOEWPTs}
\label{subsec-scan-results}

In this section, we investigate the characteristics of the parameter space that support an SFOEWPT or yield a GW signal detectable at LISA with a SNR greater than 10.
To this end, we perform a random scan over the following ranges of physical parameters:
\begin{equation}
\label{eq-scan-range}
\begin{alignedat}{3} 
\mh &\in [30,120]\GeV, &\quad
\ma &\in [30, 700]\GeV, &\quad
\mch &\in [80, 700]\GeV, \quad \\
\sba &\in [-0.5, 0.5], &\quad
\tb &\in [1, 50], &\quad
m_{12}^2 &\in [0, 2\times 10^4]\GeV^2.
\end{alignedat}
\end{equation}
Positive values for $m_{12}^2$ are exclusively scanned because the condition for avoiding a \enquote{panic vacuum}~\cite{Barroso:2013awa, Ivanov:2015nea} requires $m_{12}^2 > 0$~\cite{Dorsch:2017nza}.

For the parameter points generated within the ranges of \autoref{eq-scan-range}, we cumulatively apply three categories of constraints to define distinct sets of points for our analysis.
For clarity in the subsequent discussion, we adopt the following terminology:
\begin{itemize}
\item \textbf{Physical parameter points:} Parameter points that satisfy all the theoretical requirements and current experimental constraints.
\item \textbf{SFOEWPT parameter points:} Physical parameter points that additionally satisfy the SFOEWPT condition, $\xi_c > 1.3$.
\item \textbf{GW parameter points:} Physical parameter points that also yield a LISA GW signal with an $\text{SNR} > 10$.
\end{itemize}

Let us begin by outlining the constraints that define the physical parameter points.  
The theoretical and experimental requirements include:
\begin{enumerate}[label=(\arabic*)]
  \item vacuum stability~\cite{Ivanov:2008cxa, Barroso:2012mj, Barroso:2013awa};
  \item the condition that the Higgs potential is bounded from below~\cite{Ivanov:2006yq};
  \item tree-level unitarity in scalar–scalar scattering processes~\cite{Branco:2011iw, Arhrib:2000is};
  \item perturbativity of the Higgs quartic couplings~\cite{Chang:2015goa};
  \item consistency with the latest best-fit values for the electroweak oblique parameters ($S$, $T$, $U$)~\cite{Peskin:1991sw}, as interpreted within the 2HDM framework~\cite{He:2001tp, Grimus:2008nb}.  
  The specific values used are $S = -0.04 \pm 0.10$, $T = 0.01 \pm 0.12$, and $U = -0.01 \pm 0.09$~\cite{ParticleDataGroup:2024cfk}, with proper treatment of parameter correlations;
  \item constraints from FCNC processes at the 95\% confidence level, including bounds from $B_{d,s}\to \mu^+\mu^-$~\cite{Haller:2018nnx}, $B \to X_s \gamma$~\cite{Haller:2018nnx,Arbey:2017gmh,Sanyal:2019xcp,Misiak:2017bgg}, $B \to K^* \gamma$~\cite{Belle:2017hum}, and $B_s \to \phi \gamma$~\cite{Belle:2014sac};
  \item limits from direct searches at LEP, Tevatron, and the LHC;
  \item constraints from Higgs precision measurements.
\end{enumerate}
The theoretical conditions (1)–(5) are evaluated using the publicly available code \package{2HDMC}~\cite{Eriksson:2009ws}, while the experimental constraints (6)–(8) are implemented using \package{ScannerS} version~2~\cite{Muhlleitner:2020wwk} and \package{HiggsTools}-v1.2~\cite{Bahl:2022igd}.

We briefly comment on the treatment of the Higgs precision measurement constraints.  
For each 2HDM parameter point that satisfies constraints (1) through (7), we compute the global minimum of the Higgs precision $\chi^2$, denoted by $\chi^2_{\min}(\text{2HDM})$.  
We then evaluate, for every point, the difference $\Delta\chi^2 = \chi^2 - \chi^2_{\min}(\text{2HDM})$.  
Given that the model contains six free parameters, as listed in \autoref{eq-model-parameters}, we accept parameter points with $\Delta\chi^2 < 12.59$, corresponding to the 95\% confidence level for six degrees of freedom.

From this comprehensive scan and filtering, we obtained $2.36\times 10^6$ physical parameter points.
For these points, we then perform a detailed analysis of the thermal phase transition history.
This includes calculations of the one-loop effective potential at finite temperature, the critical temperature $T_c$, its corresponding VEV $v_c$, the nucleation temperature $T_n$ (obtained by solving the bounce equation), the percolation temperature $T_p$, and the resulting GW spectrum.
Given the complexity of these calculations, several public packages are available for numerical analysis, such as \package{CosmoTransitions}~\cite{Wainwright:2011kj}, \package{PhaseTracer}~\cite{Athron:2020sbe}, and \package{BSMPT}~\cite{Basler:2018cwe,Basler:2020nrq,Basler:2024aaf}.
We choose \package{BSMPT} version 3.0.7 for our numerical analysis, primarily because it is written in \texttt{C++}, which significantly reduces computation time.
Moreover, it is adept at analyzing multi-step EWPTs by tracing multiple temperature-dependent vacuum phases across extended Higgs potentials.
The code automatically identifies and characterizes each transition step in the sequence---including associated critical, nucleation, and percolation temperatures---making it particularly powerful for studying complex transition histories involving several vacuum directions.

In our analysis, we used the executable \texttt{CalcGW} to determine the critical temperature $T_c$, nucleation temperature $T_n$, percolation temperature $T_p$, and vacuum expectation values (VEVs) at $T_c$ for each transition step. Additionally, we evaluated GW SNRs at LISA by setting $T_* = T_p$ and adopting a default bubble wall velocity $\vwl = 0.95$.

We retained only parameter points satisfying the following criteria: (i) the diagnostic output \texttt{status\_ewsr} equals \texttt{ew\_sym\_res}; and (ii) successful (\texttt{success}) diagnostics for \texttt{status\_nlo\_stability},  \texttt{status\_tracing}, \texttt{status\_coex\_pairs}, \texttt{status\_crit},  
\texttt{status\_bounce\_sol}, \texttt{status\_nucl}, \texttt{status\_perc}, \texttt{status\_compl}, and \texttt{status\_gw}. For these points, we subsequently reran the executable \texttt{MinimaTracer} to verify the absence of vacuum trapping~\cite{Biekotter:2022kgf,Chen:2025ksr}. Vacuum trapping is a critical issue in FOEWPT scenarios, in which the Universe remains stuck in a metastable vacuum despite the existence of a deeper true vacuum with $v \approx 246\GeV$ at zero temperature.
 Approximately 5\% of the identified SFOEWPT parameter points were found to suffer from vacuum trapping.

\begin{figure}[t]
\centering
\includegraphics[width=\textwidth]{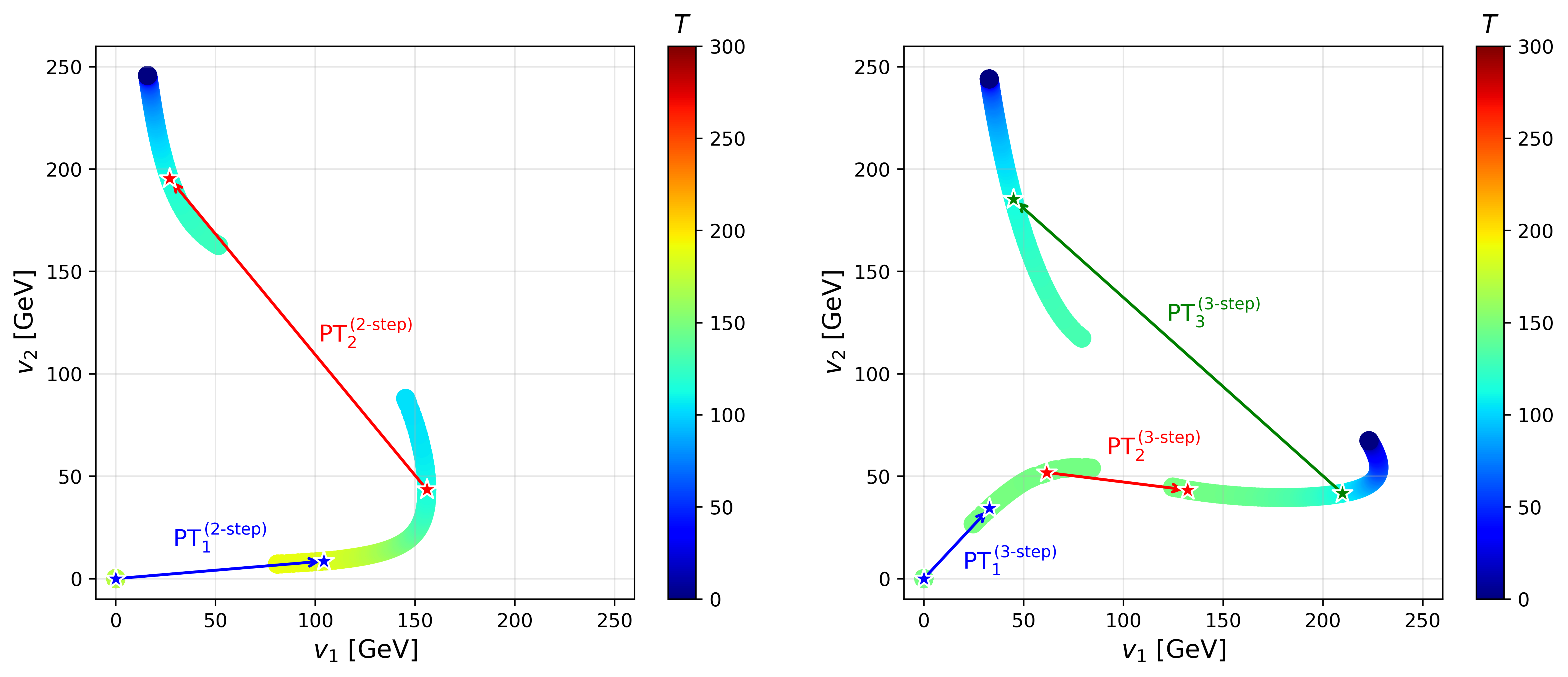}
\caption{%
	The evolution of the minima of an effective potential with two relevant scalar field directions, as a function of temperature in units of GeV (indicated by the color scale).
	The left panel exemplifies a two-step EWPT for the benchmark point $\text{BP}_\text{2-step}$, while the right panel presents an example of a three-step EWPT for $\text{BP}_\text{3-step}$.
	Both benchmark points are described in the main text.
}
\label{fig-evolution} 
\end{figure}

We begin by investigating whether the inverted Type-I 2HDM can support an SFOEWPT and, if so, whether it proceeds via a one-step, two-step, or higher-order transition pathway.
To systematically label individual transitions within an $n$-step EWPT, we define the notation $\trans{i}{n}$ as:
\[
\trans{i}{n}:\thickspace\text{the $i$-th transition in an $n$-step FOEWPT}.
\]

To show the evolution through sequential vacuum states in such multi-step thermal transitions, \autoref{fig-evolution} presents representative examples of the VEV evolution in the effective potential (projected onto two relevant field directions) for two-step (left panel) and three-step (right panel) EWPTs.
In these plots, the color scale indicates the temperature.
These illustrative multi-step EWPTs correspond to the benchmark points $\text{BP}_\text{2-step}$ and $\text{BP}_\text{3-step}$, which are defined by the following parameters:
\begin{equation}
\label{eq-2-3step}
\begin{alignedat}{3}
\text{BP}_\text{2-step}: \quad
	\mh &= 83.4\GeV, &\quad
	\ma &= 194.5\GeV, &\quad
	\mch &= 178.9\GeV, \\
    m_{12}^2 &= 424.9\GeV^2, &\quad
    \tb &= 15.4, &\quad
    \sba &= 0.114; \\
\text{BP}_\text{3-step}: \quad
    \mh &= 88.2\GeV, &\quad
    \ma &= 155.4\GeV, &\quad
    \mch &= 184.3\GeV, \\
    m_{12}^2 &= 1001.7\GeV^2, &\quad
    \tb &= 7.4, &\quad
    \sba &= 0.212.
\end{alignedat}
\end{equation}

For transitions originating from a metastable (non-symmetric) vacuum in a multi-step sequence, the definitions of the critical temperature in \autoref{eq-Tc-1step} and the order parameter $\xi_c$ in \autoref{eq-xic-one-step} must be generalized.  
Consider the $i$-th transition in an $n$-step sequence, $\trans{i}{n}$.  
Prior to this transition—but following the completion of the preceding step $\trans{i-1}{n}$—the Universe is assumed to reside in a global minimum of the effective potential at $\bigl(v_{1(i-)}^T, v_{2(i-)}^T\bigr)$.
Note that we primarily focus on the dynamics within an effective two-field configuration, as the magnitude of $v_3^T$ remains consistently negligible compared to $v_1^T$ and $v_2^T$ across the relevant temperature range (typically suppressed by factors of $10^{-6}$ or more).  
As the Universe continues to cool, another local minimum at $\bigl(v_{1(i+)}^T, v_{2(i+)}^T\bigr)$ may appear and become dynamically relevant.

When the temperature decreases further, these two  minima can achieve equal effective potential values.
This condition defines the critical temperature $T_c$ for the specific transition $\trans{i}{n}$:
\beq
\label{eq-Tc-multi} 
\veff\bigl(v_{1(i-)}^{T_c},v_{2(i-)}^{T_c}, T_c\bigr) = \veff\bigl(v_{1(i+)}^{T_c},v_{2(i+)}^{T_c}, T_c\bigr).
\eeq
The strength of this $\trans{i}{n}$ transition is then characterized by the order parameter $\xi_c$:
\beq
\label{eq-xic-multi} 
\xi_c = \frac{\sqrt{\bigl(v_{1(i+)}^{T_c} - v_{1(i-)}^{T_c}\bigr)^2 + \bigl(v_{2(i+)}^{T_c} - v_{2(i-)}^{T_c}\bigr)^2}}{T_c},
\eeq
which quantifies the magnitude of the VEV change relative to the critical temperature for that particular step.

It is important to note that both $T_c$ and $\xi_c$ can vary significantly across the individual transitions $\trans{i}{n}$ within a multi-step sequence.  
This feature is illustrated by the critical temperatures and order parameters associated with each transition stage for the benchmark points given in \autoref{eq-2-3step}:
\bea
\label{eq-2-3step-xic-Tc}
\begin{alignedat}{2}
\text{BP}_\text{2-step}: \quad
	T_c &= 186.3\GeV, &\quad \xi_c &= 0.562, \quad\text{for $\trans{1}{2}$}; \\
	T_c &= 115.2\GeV, &\quad \xi_c &= 1.731, \quad\text{for $\trans{2}{2}$}; \\[10pt]
\text{BP}_\text{3-step}: \quad
	T_c &= 148.3\GeV, &\quad \xi_c &= 0.321, \quad\text{for $\trans{1}{3}$}; \\
	T_c &= 147.4\GeV, &\quad \xi_c &= 0.483, \quad\text{for $\trans{2}{3}$}; \\  
	T_c &= 113.2\GeV, &\quad \xi_c &= 1.930, \quad\text{for $\trans{3}{3}$}.
\end{alignedat}
\eea
In both examples, the largest $\xi_c$ occurs at the final stage of the transition sequence. This highlights that focusing solely on the first transition from the symmetric phase may overlook cases where an SFOEWPT emerges only at a later stage in a multi-step evolution.

A central question is how many parameter points support an SFOEWPT, and more specifically, which transition step within a multi-step EWPT is most likely to be strong.  
Out of the $2.36 \times 10^6$ physically viable parameter points identified in our scan,
the number of points that induce an SFOEWPT via a given transition step $\trans{i}{n}$ is summarized below:\footnote{%
	These results depend on the specific algorithms used to trace the VEVs and identify minima of the effective potential at finite temperature. We employed the default mode in the \texttt{Multi-Step Phase Transition} module of \package{BSMPT}, which relies on the \texttt{minimum\_tracer}. Other modes or algorithms may yield different results.
}
\bea
\label{eq-probability}
\nsfoewpt^{\trans{1}{1}} = 5343, \quad
\nsfoewpt^{\trans{1}{2}} = 200,   \quad \nsfoewpt^{\trans{2}{2}} = 4486, \quad
\nsfoewpt^{\trans{3}{3}} = 6. 
\eea
All other transition stages, including transitions involving four or more steps, yield zero instances of an SFOEWPT.  
An interesting finding is that out of the 200 parameter points contributing to $\nsfoewpt^{\trans{1}{2}}$, 186 also induce an SFOEWPT via $\trans{2}{2}$, thereby realizing two successive SFOEWPTs.  
This implies that only 14 parameter points exhibit an SFOEWPT exclusively through $\trans{1}{2}$.

Overall, approximately 0.42\% of all physical parameter points accommodate an SFOEWPT in at least one transition step.  
The largest fraction arises from $\trans{1}{1}$, followed closely by $\trans{2}{2}$, which exhibits a comparable probability for supporting an SFOEWPT.  
Although much rarer, the model can also yield an SFOEWPT in a three-step sequence, predominantly at the final step $\trans{3}{3}$.

Although the numbers of parameter points presented in \autoref{eq-probability} are small relative to the total number of physical parameter points, they clearly indicate that a non-negligible region of the parameter space in the inverted Type-I 2HDM can support an SFOEWPT.  
Moreover, it is important to emphasize that Nature selects only a single parameter point; thus, the overall small percentage does not, by itself, disfavor the model as a viable framework for realizing an SFOEWPT.

\subsection{Characteristics of SFOEWPT Parameter Points in Multi-step Transitions}
\label{subsec-SFOEWPT-characteristics}

\begin{figure}[t]
\centering
\includegraphics[width=\textwidth]{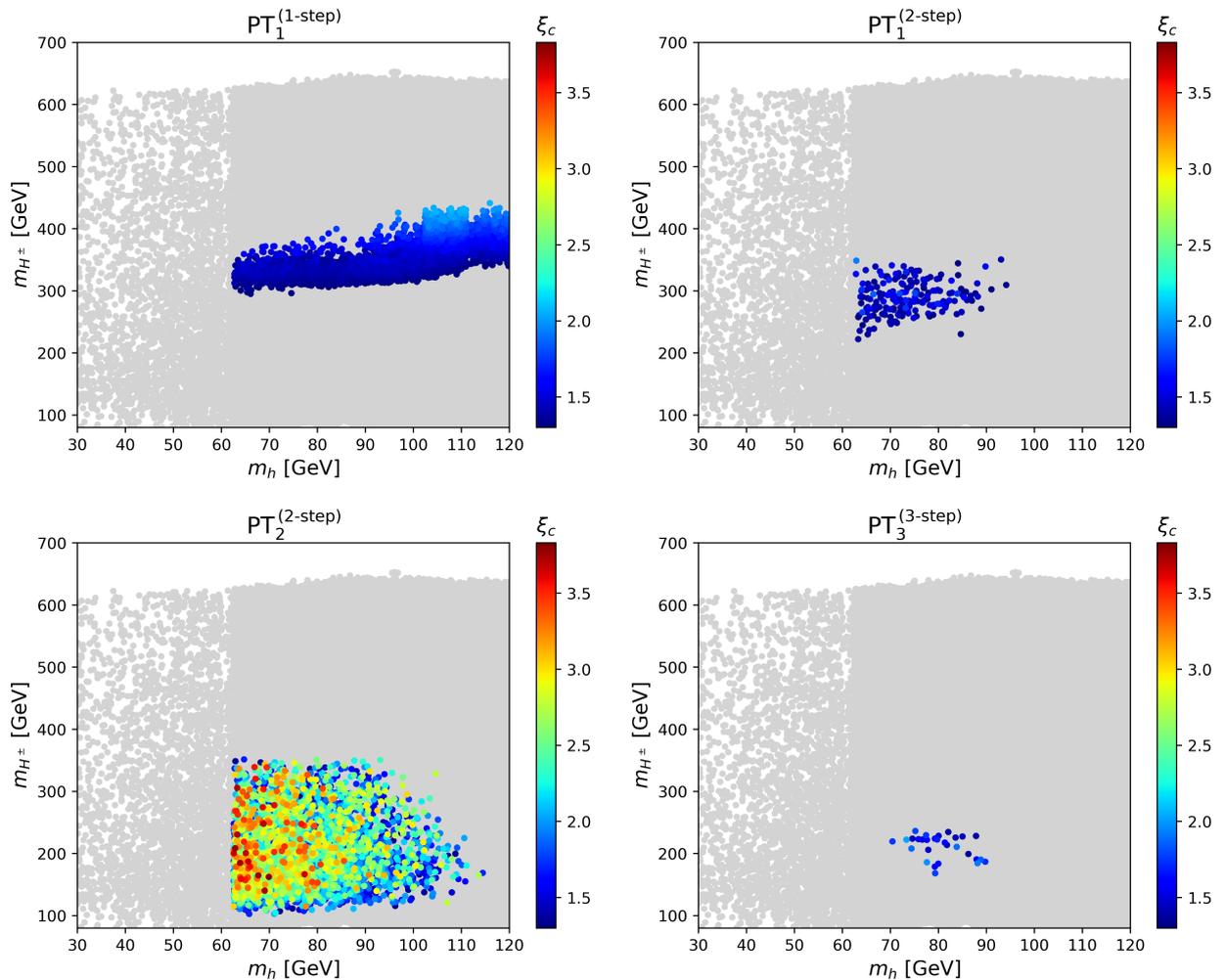} 
\caption{%
	$\mch$ vs $\mh$ for SFOEWPT parameter points ($\xi_c > 1.3$), with $\xi_c$ indicated by the color scale.
	Grey points represent all physical parameter points satisfying theoretical and current experimental constraints.
	The panels display results for $\trans{1}{1}$ (upper left), $\trans{1}{2}$ (upper right), $\trans{2}{2}$ (lower left), and $\trans{3}{3}$ (lower right).
}
\label{fig-mch-mh-xic} 
\end{figure}

This subsection investigates the properties of SFOEWPT parameter points.  
As a primary illustration, \autoref{fig-mch-mh-xic} shows the distribution of these points in the $(\mh, \mch)$ plane, color-coded by the order parameter $\xi_c$, across various multi-step transition types:  
$\trans{1}{1}$ in the upper-left panel;  
$\trans{1}{2}$ in the upper-right panel;  
$\trans{2}{2}$ in the lower-left panel;  
and $\trans{3}{3}$ in the lower-right panel.
In each panel, points are layered by $\xi_c$, with higher values plotted on top to emphasize the stronger transitions.

Note that the results for $\trans{3}{3}$ are based on a total of 29 parameter points, rather than the smaller sample reported in \autoref{eq-probability}.
Because SFOEWPTs via $\trans{3}{3}$ were too rare in the general scan (with only six points identified), we performed an additional targeted scan over a narrower parameter range:
$\mh \in [70,90]\GeV$, $\ma \in [110,250]\GeV$, $\mch \in [160,240]\GeV$, $\sba \in [0,0.25]$, $\tan\beta \in [3,8]$, and $m_{12}^2 \in [700,1300]\GeV^2$, generating $4\times10^5$ additional physical parameter points.
This yielded 23 new SFOEWPT points via $\trans{3}{3}$, bringing the total to 29 for this transition.
All of these are included in the lower-right panel of \autoref{fig-mch-mh-xic}.

In \autoref{fig-mch-mh-xic}, we also show, for comparison, the full set of physical parameter points as underlying gray scatter points.  
These illustrate how theoretical requirements and current experimental data constrain the parameter space in the inverted Type-I 2HDM.
In particular, the mass of the charged Higgs boson $H^\pm$ is bounded from above, with $\mch \lesssim 630\GeV$.  
It is also notable that the lighter \textit{CP}-even Higgs boson $h$ can have a mass below $\mhsm/2$, although this region is sparsely populated compared to the region where $\mh > \mhsm/2$.  
Such points remain viable due to the possibility of a \enquote{Higgs-phobic} $h$, where the $H$–$h$–$h$ coupling is sufficiently suppressed to evade stringent constraints from null results in $H_\text{SM} \to hh$ searches~\cite{Kim:2022hvh}.

The additional requirement of an SFOEWPT imposes further stringent constraints on the allowed regions for $\mch$ and $\mh$.  
A common feature across all transition types is the exclusion of parameter points with $\mh < \mhsm/2$ by the SFOEWPT condition.\footnote{%
	Relaxing the SFOEWPT criterion—for example, to $\xi_c > 1$—does not significantly alter this outcome.
}  
Beyond this shared feature, the specific impact of the SFOEWPT requirement on $\mch$ and $\mh$ depends on the transition path.  
For one-step transitions, $\mch$ is tightly constrained to the range $[295, 441]\GeV$, while $\mh$ spans $[62.5, 120]\GeV$.  
In contrast, $\trans{2}{2}$ transitions—which make up the second-largest class of SFOEWPT parameter points—permit a broader range, with $\mch \in [100, 350]\GeV$.  
For the remaining $\trans{1}{2}$ and $\trans{3}{3}$ transitions, the SFOEWPT condition imposes even tighter restrictions on both $\mch$ and $\mh$.

These results---particularly those concerning the allowed ranges for $\mch$---provide significant insight into our primary question: \textit{are the parameter regions that induce a SFOEWPT via one-step and multi-step transitions distinct?}  The answer, based on our scan, is that they are largely distinct, though a narrow region of overlap exists. To illustrate this, let us focus on the two dominant transition types, $\trans{1}{1}$ and $\trans{2}{2}$. If a charged Higgs boson were observed with a mass $\mch > 351\GeV$, our findings imply that any associated SFOEWPT must have proceeded via a one-step transition. Conversely, if an observation yielded $\mch < 295\GeV$, only a two-step transition could have facilitated the SFOEWPT. In the intermediate mass window, $295\GeV \lesssim \mch \lesssim 351\GeV$, both transition types remain possible; an SFOEWPT in this range could arise from either a $\trans{1}{1}$ or a $\trans{2}{2}$ pathway. This overlap, however, is relatively limited compared to the full mass ranges allowed for each transition type, underscoring a partial but meaningful separation in their respective parameter spaces.

Another notable feature revealed in \autoref{fig-mch-mh-xic} is that $\xi_c$ varies significantly depending on the specific stage of the multi-step transition sequence.  
Among the transition types considered, $\trans{2}{2}$ stands out by supporting the strongest SFOEWPTs, with $\xi_c$ values reaching up to approximately 3.83.  
Moreover, for this transition type, parameter points with larger $\xi_c$ values tend to cluster in regions where the lighter \textit{CP}-even Higgs boson $h$ is also relatively light.  
In contrast, the other transition stages—$\trans{1}{1}$, $\trans{1}{2}$, and $\trans{3}{3}$—typically yield more moderate phase transition strengths, with $\xi_c$ values generally below 2.08.

\begin{figure}[t]
\centering
\includegraphics[width=\textwidth]{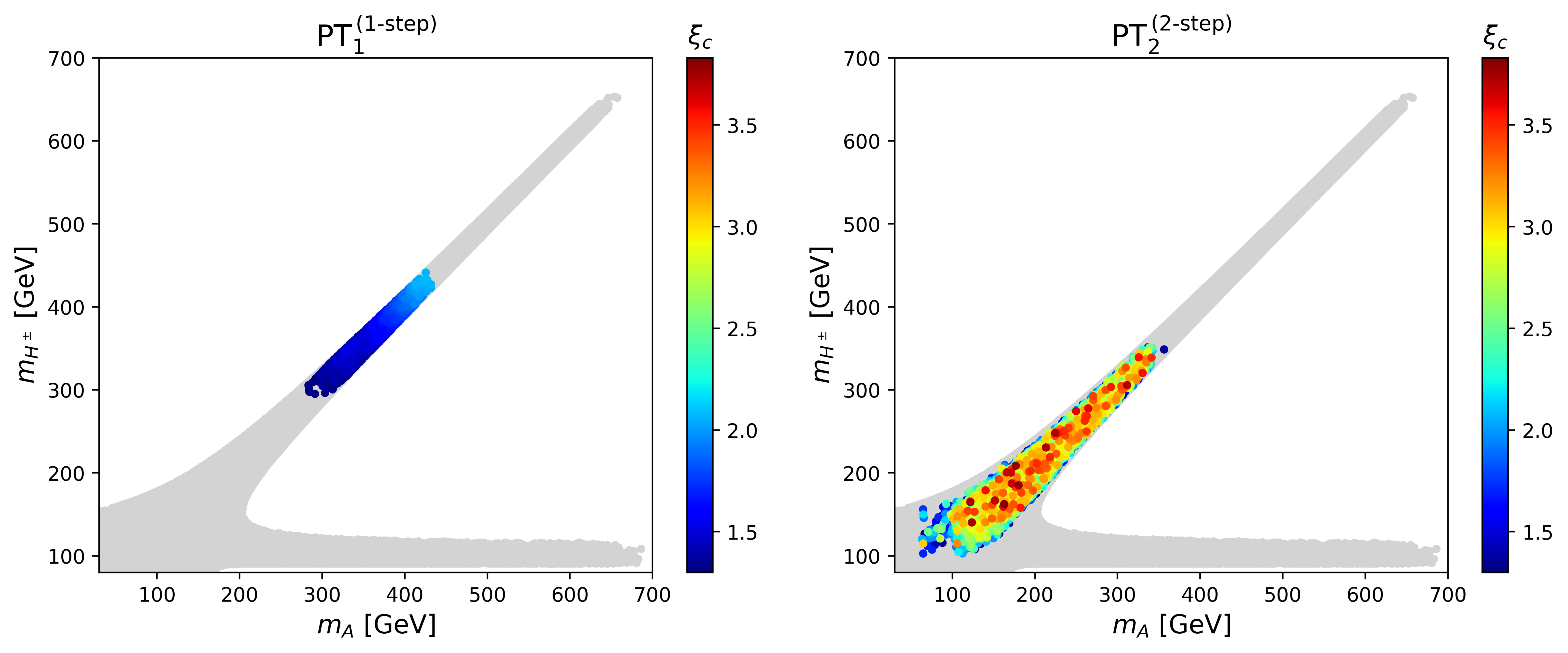} 
\caption{%
	$\mch$ vs $\ma$ for SFOEWPT parameter points with $\xi_c$ indicated by the color scale.
	Grey points represent all physical parameter points.
	The left panel shows results for $\trans{1}{1}$ and the right panel for $\trans{2}{2}$.
}
\label{fig-mch-ma-xic}
\end{figure}

To investigate the characteristics of other model parameters, we now focus on two primary stages which can induce SFOEWPT, $\trans{1}{1}$ and $\trans{2}{2}$.
\autoref{fig-mch-ma-xic} presents scatter plots of $\mch$ versus $\ma$ for SFOEWPT parameter points, color-coded by the corresponding order parameter $\xi_c$.
The left panel displays results for $\trans{1}{1}$ and the right panel for $\trans{2}{2}$, with grey points again representing all physical parameter points.

The distribution of physical parameter points (grey) reveals two distinct regions: one where $\mch \sim \ma$ and another where $\mch \sim \mh$.
This structure is primarily attributed to constraints from the Peskin-Takeuchi oblique parameters ($S, T, U$), as BSM Higgs contributions to these parameters tend to be suppressed when at least two of the new scalar masses are degenerate.
However, the requirement of an SFOEWPT effectively excludes the second region ($\mch \sim \mh$), favoring a mass degeneracy between $\mch$ and $\ma$, for both $\trans{1}{1}$ and $\trans{2}{2}$ transitions.
Nevertheless, the allowed mass ranges for $\mch$ and $ \ma$ vary significantly depending on the transition stage.
An SFOEWPT via $\trans{1}{1}$ typically requires relatively heavy charged Higgs bosons and pseudoscalars: $\mch \in [295,441]\GeV$ and $\ma \in [283,432]\GeV$.
In contrast, an SFOEWPT via $\trans{2}{2}$ permits lighter $H^\pm$ and $A$: $\mch \in [103,351]\GeV$ and $\ma \in [63,356]\GeV$.
Notably, for $\trans{2}{2}$ transitions, it is feasible for the pseudoscalar $A$ to be lighter than the $125\GeV$ SM-like Higgs boson.
Consequently, future observations of $\mch$ and $\ma$ could help distinguish whether an SFOEWPT, if one indeed occurred, proceeded via a one-step or two-step EWPT, except within the narrow overlap region of $\mch \in [295,351]\GeV$ and $\ma \in [283,356]\GeV$.

\begin{figure}[t]
\centering
\includegraphics[width=\textwidth]{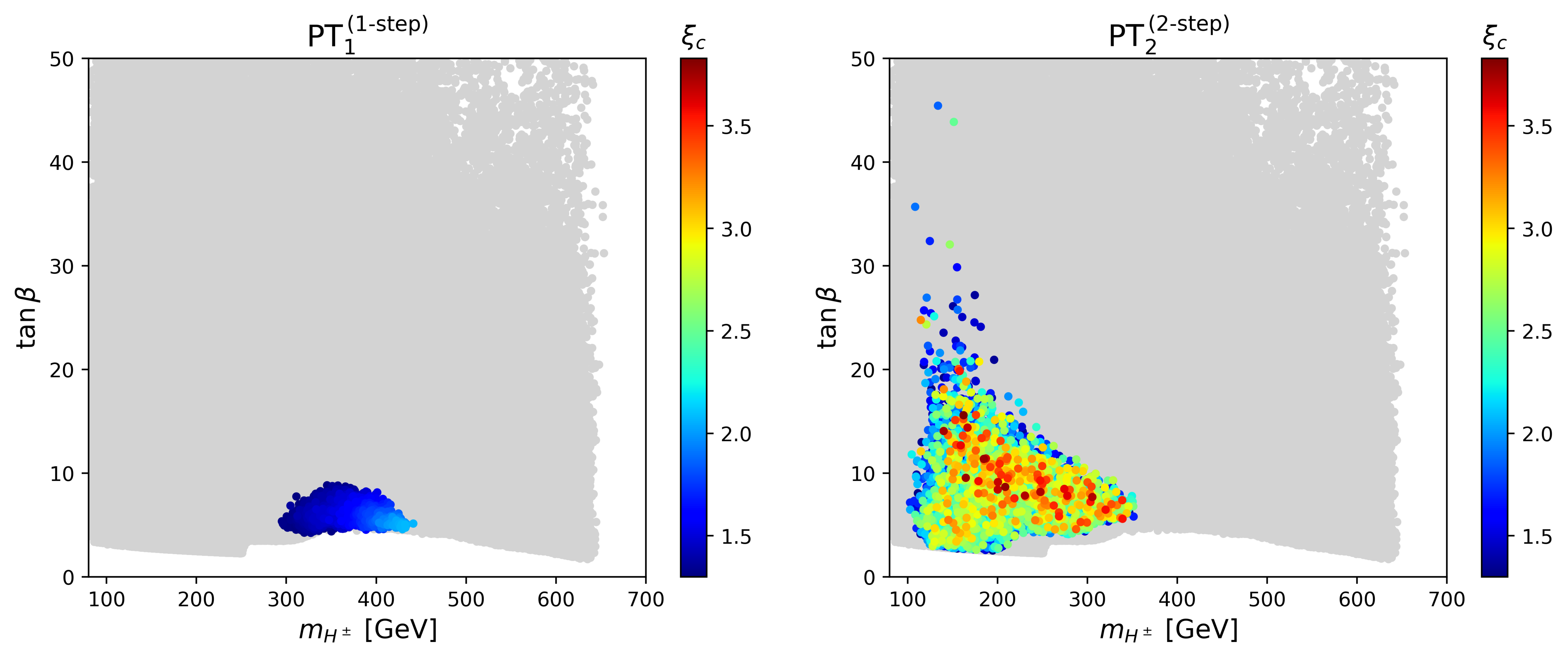} 
\caption{%
	$\tb$ versus $\mch$ for SFOEWPT parameter points, with the color code of $\xi_c $.
	Grey points represent all physical parameter points.
	The left panel shows results for $\trans{1}{1}$ and the right panel for $\trans{2}{2}$.
}
\label{fig-tb-mch-xic}
\end{figure}

\autoref{fig-tb-mch-xic} shows the distribution of $\tb$ versus $\mch$ for SFOEWPT parameter points, with $\xi_c$ indicated by the color scale, for the $\trans{1}{1}$ (left) and $\trans{2}{2}$ (right) transitions.
For the full set of physical parameter points (shown in grey), the allowed range $\tb \in [1.72, 50]$ covers most of the scanned domain.
The region $\tb \in [1, 1.72]$ is excluded, primarily due to experimental constraints from FCNC measurements, as lower $\tb$ values enhance the charged Higgs Yukawa couplings and lead to sizable contributions to FCNC observables.
In contrast, large $\tb$ values remain allowed, as they suppress the BSM Higgs Yukawa couplings (see \autoref{eq-Yukawa}), thereby reducing their experimental impact—a well-known feature of the Type-I 2HDM.

Interestingly, the SFOEWPT condition imposes a stringent constraint on $\tb$.  
For $\trans{1}{1}$ transitions, only a narrow band, $\tb \in [4.23, 8.82]$, remains viable.  
In contrast, $\trans{2}{2}$ transitions accommodate a much broader range, with $\tb$ values extending up to 45.3.  
However, high-$\tb$ points (e.g., $\tb > 30$) appear with lower density and are confined to regions with relatively light charged Higgs bosons.  
A mild correlation between $\tb$ and $\mch$ is also observed in the $\trans{2}{2}$ case, where heavier $H^\pm$ masses are generally associated with smaller $\tb$ values.  
Although the EWPT strength $\xi_c$ does not show a strong direct dependence on $\tb$, points with $\xi_c > 3$ are predominantly located in the region $\tb \lesssim 20$.

\begin{figure}[t]
\centering
\includegraphics[width=\textwidth]{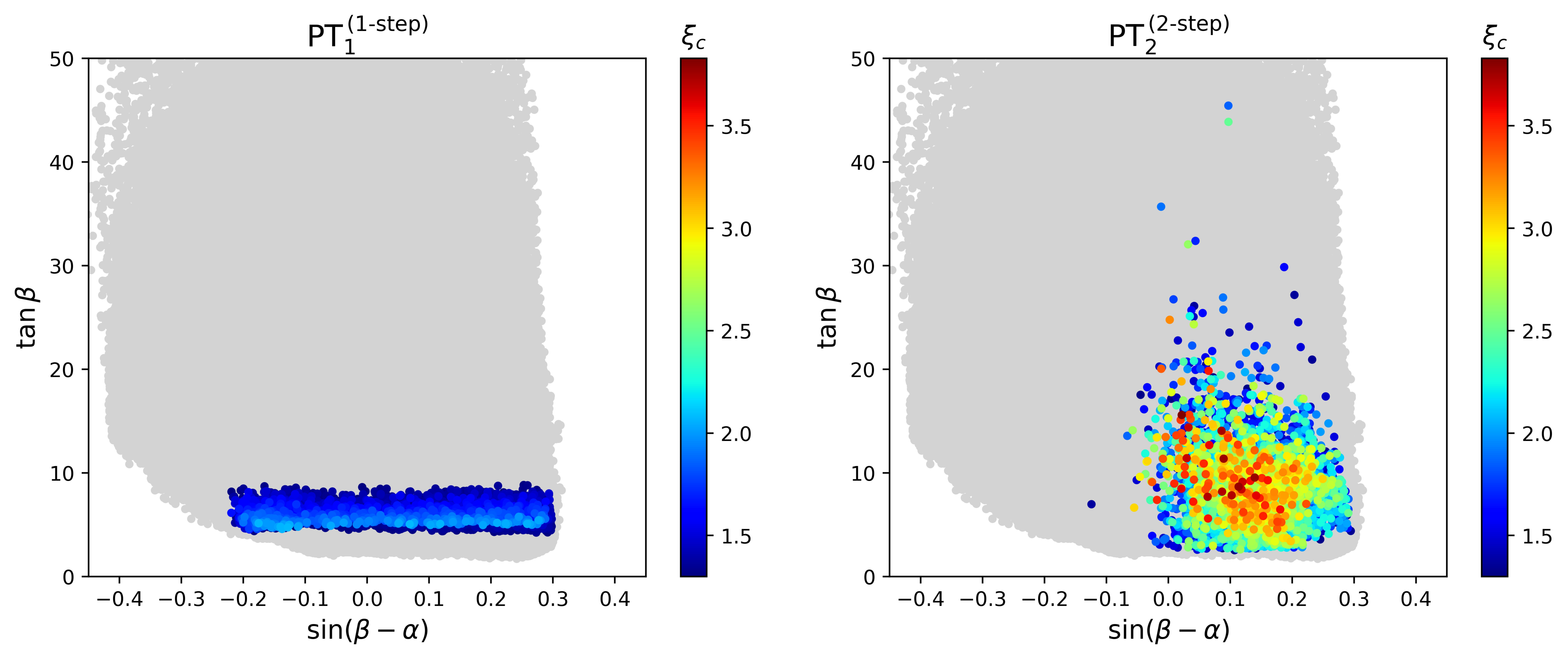} 
\caption{%
	$\tb$ versus $\sba$ for SFOEWPT parameter points, with $\xi_c (>1.3)$ indicated by the color scale.
	Grey points represent all physical parameter points.
	The left panel shows results for $\trans{1}{1}$ and the right panel for $\trans{2}{2}$.
}
\label{fig-tb-sba-xic}
\end{figure}

We now examine the characteristics of $\tb$ versus $\sba$ for SFOEWPT parameter points, as illustrated in \autoref{fig-tb-sba-xic}.
The left panel shows results for the $\trans{1}{1}$ transition, and the right panel for $\trans{2}{2}$, with color codes indicating the corresponding $\xi_c$ values.
Grey points represent the full set of physical parameter points.
In the Type-I 2HDM, sizable deviations from the Higgs alignment limit are permitted by current theoretical and experimental constraints, with $\sba$ spanning the range $[-0.460, 0.314]$.

Imposing the SFOEWPT condition substantially narrows the allowed range of $\sba$.
Most notably, the preferred values differ significantly between $\trans{1}{1}$ and $\trans{2}{2}$ transitions.  
For $\trans{1}{1}$ transitions, the viable range is reduced to $\sba \in [-0.22, 0.30]$, with negative values comprising approximately 44\% of the SFOEWPT parameter points.  
Although more constrained, sizable deviations from the alignment limit remain allowed.  
This behavior in one-step SFOEWPTs is closely tied to the vacuum uplifting measure $\dfz$.  
In such transitions, larger $\xi_c$ values correlate with more positive $\dfz$ (see \autoref{eq-dfz-onestep}).  
In the inverted scenario, the tree-level contribution $\dfz^{\text{tree}}$ (see \autoref{eq-dfz-tree}) favors nonzero $\sba$ to enhance $\dfz$, thus promoting departures from alignment.  
However, $|\sba|$ cannot be arbitrarily large: excessively positive $\dfz^{\text{tree}}$ pushes the total $\dfz$ beyond the vacuum stability threshold $\fzasm$, leading to an unwanted metastable vacuum at zero temperature.  
One-loop contributions to $\dfz$ in the alignment limit (\autoref{eq-dfz-sba0}) are also strictly positive and reinforce this behavior.

In contrast, SFOEWPTs proceeding via $\trans{2}{2}$ exhibit a strong preference for positive $\sba$ values.  
As shown in the right panel of \autoref{fig-tb-sba-xic}, negative $\sba$ values make up only about 1\% of the SFOEWPT parameter points for $\trans{2}{2}$.  
Although not explicitly shown, similar trends are observed for $\trans{1}{2}$ and $\trans{3}{3}$ transitions, where negative $\sba$ values comprise approximately 3\% and 0\% of the respective SFOEWPT samples.

This yields a valuable phenomenological insight: if future measurements or global fits favor a negative value of $\sba$, then an SFOEWPT—if it indeed occurred—likely proceeded via a one-step transition.
However, the converse does not hold: a one-step SFOEWPT does not imply negative $\sba$, as $\trans{1}{1}$ transitions also accommodate positive values of $\sba$.

\begin{figure}[t]
\centering
\includegraphics[width=\textwidth]{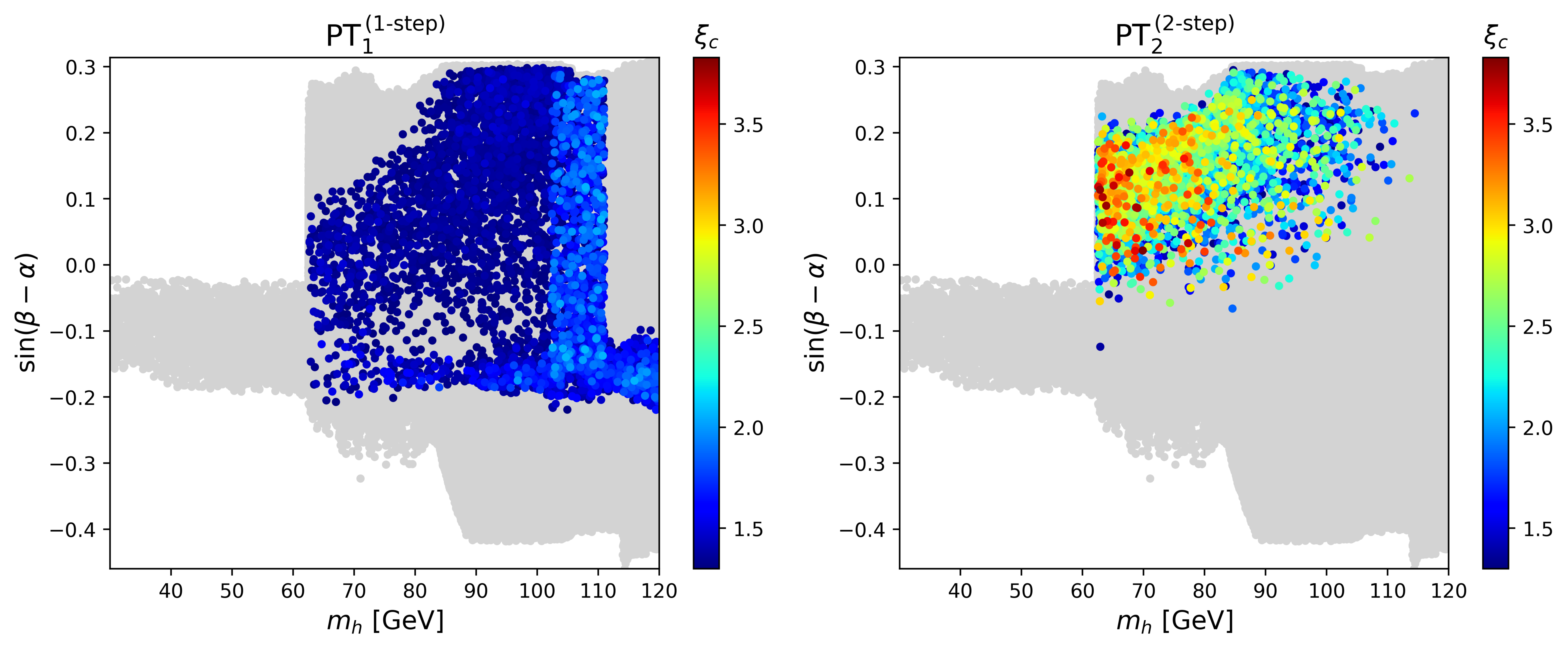} 
\caption{%
	$\sba$ versus $\mh$ for SFOEWPT parameter points, with $\xi_c$ indicated by the color scale.
	Grey points represent all physical parameter points.
	The left panel shows results for $\trans{1}{1}$ and the right panel for $\trans{2}{2}$.
}
\label{fig-sba-mh-xic}
\end{figure}

Finally, we present an interesting correlation between $\sba$ and $\mh$ for SFOEWPT points in \autoref{fig-sba-mh-xic}, with $\xi_c$ as the color code.
Grey points again denote all physical parameter points.
The left panel shows results for $\trans{1}{1}$ and the right panel for $\trans{2}{2}$.
For $\trans{1}{1}$, as $\mh$ increases from approximately $62.5\GeV$ up to around $85\GeV$, the allowed range of $\sba$ broadens.
For $\mh \in [85,100]\GeV$, nearly the entire SFOEWPT-allowed range of $\sba \in [-0.22, 0.30]$ is populated.
More strikingly, for $\mh$ in the $[110,120]\GeV$ range, only negative values, $\sba \in [-0.22,-0.10]$ are found to accommodate SFOEWPT.
This implies that if $h$ is relatively heavy, an SFOEWPT via a one-step transition in this model requires a sizable deviation from the Higgs alignment limit, specifically towards negative $\sba$.

For SFOEWPTs via $\trans{2}{2}$, the distribution of parameter points in the $\sba$ versus $\mh$ plane exhibits a relatively simple structure, with points clustering in a compact region without any notable features or peculiar shapes.  
An interesting observation is that for $\mh \gtrsim 100\GeV$, the Higgs alignment limit is no longer maintained in this case.

Let us summarize the allowed regions of the SFOEWPT parameter space for the $\trans{1}{1}$ and $\trans{2}{2}$ transitions:
\begin{equation}
\label{eq-summary-param}
\begin{alignedat}{2}
\text{SFOEWPT via $\trans{1}{1}$:} \quad \xi_c &\in [1.30, 2.08], & & \\
	\mh &\in [62.7, 120.0]\GeV, &\quad \ma &\in [283.4, 431.9]\GeV, \\
	\mch &\in [295.1, 441.3]\GeV, &\quad m_{12}^2 &\in [0.077, 2549]\GeV^2, \\
	\tb &\in [4.2, 8.8], &\quad \sba &\in [-0.221, 0.30]; \\[10pt]
\text{SFOEWPT via $\trans{2}{2}$:} \quad \xi_c &\in [1.30, 3.83], & & \\
	\mh &\in [62.5, 114.4]\GeV, &\quad \ma &\in [63.2, 356.4]\GeV, \\
	\mch &\in [102.7, 351.4]\GeV, &\quad m_{12}^2 &\in [29.0, 2775]\GeV^2, \\
	\tb &\in [2.5, 45.4], &\quad \sba &\in [-0.12, 0.30].
\end{alignedat}
\end{equation}
We emphasize that the SFOEWPT parameter points are not uniformly distributed across these ranges, as evident from the scatter plots presented in this subsection.

\subsection{Correlation between Vacuum Uplifting and $\xi_c$ in Multi-step Transitions}
\label{subsec-correlation-DF0-xic}

In \autoref{subsec-DF0-general}, we discussed a well-established feature of SFOEWPTs in one-step transitions within the normal Higgs scenario: the vacuum uplifting measure $\dfz$, which represents the contrast between the 2HDM and SM in terms of the vacuum energy density difference between symmetric and broken phases at zero temperature, is strongly correlated with the FOEWPT strength $\xi_c$.
In this section, we examine whether such a correlation persists in the inverted Higgs scenario, particularly in the context of multi-step EWPTs.

\begin{figure}[t]
\centering
\includegraphics[width=\textwidth]{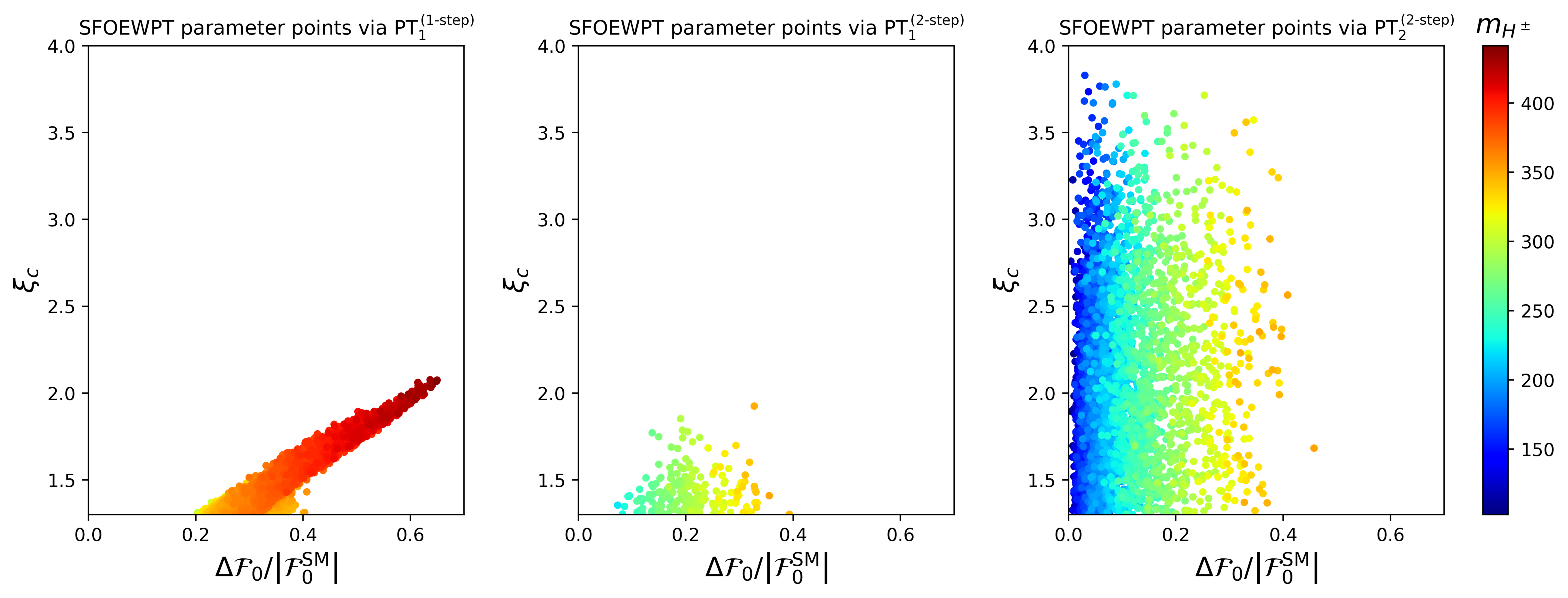} 
\caption{%
	$\xi_c$ versus $\dfz/\fzasm$ for SFOEWPT parameter points with the color code of $\mch$ in units of GeV.
	The results of $\trans{1}{1}$, $\trans{1}{2}$, and $\trans{2}{2}$ are presented in the left, middle, and right panel, respectively.
}
\label{fig-xic-DF0-mch}
\end{figure}

In \autoref{fig-xic-DF0-mch}, we show the correlation between $\xi_c$ and $\dfz/\fzasm$ for SFOEWPT parameter points, with $\mch$ indicated by the color scale.
The results for the $\trans{1}{1}$, $\trans{1}{2}$, and $\trans{2}{2}$ transitions are displayed in the left, middle, and right panels, respectively.

For the one-step transition, a clear positive correlation is observed: larger values of $\dfz$ generally correspond to stronger phase transitions, i.e., larger $\xi_c$.
In the inverted Type-I scenario, the requirement of $\xi_c > 1.3$ constrains the vacuum uplifting measure to the approximate range of
\[
\text{SFOEWPT via one-step with $\mhh = 125\gev$:} \quad 0.20 \lesssim \frac{\dfz}{\fzasm} \lesssim 0.65.
\]
This result indicates that the degree of vacuum uplifting required to realize an SFOEWPT in the inverted scenario differs from that in the normal scenario (cf.~\autoref{eq-dfz-onestep}).
In particular, the allowance of smaller $\dfz/\fzasm$ values suggests that even a 2HDM vacuum with a potential depth relatively close to that of the SM vacuum can still support an SFOEWPT.

However, the direct correlation between $\xi_c$ and $\dfz/\fzasm$ largely disappears in multi-step EWPTs.
This behavior is illustrated in the middle and right panels of \autoref{fig-xic-DF0-mch}, corresponding to the $\trans{1}{2}$ and $\trans{2}{2}$ transitions, respectively.
For both cases, no clear correlation is observed: for a fixed value of $\dfz/\fzasm$, the order parameter $\xi_c$ can span a wide range while still satisfying the SFOEWPT condition.
This lack of correlation is particularly pronounced in the $\trans{2}{2}$ transition.

Instead, a different pattern emerges in the multi-step scenarios: a correlation between $\dfz/\fzasm$ and $\mch$, where larger $\dfz/\fzasm$ values tend to be associated with heavier $\mch$.
This trend can be partly understood from the analytic structure of $\dfz$ in the Higgs alignment limit, as discussed in connection with \autoref{eq-dfz-sba0}.

In summary, our investigation of the vacuum uplifting measure $\dfz$ and its correlation to the EWPT strength $\xi_c$ in the inverted Higgs scenario reveals qualitatively distinct behaviors between one-step and multi-step transitions.
For one-step transitions, a positive correlation  is confirmed, although the viable range of $\dfz/\fzasm$ differs from that in the normal Higgs scenario.
In contrast, for multi-step transitions, this correlation breaks down entirely; the strength of the phase transition $\xi_c$ shows no clear dependence on $\dfz/\fzasm$.
As a result, in multi-step scenarios, evaluating the strength of the FOEWPT requires a full computation of the one-loop effective potential at finite temperature to determine the critical temperature and the corresponding order parameter.

\subsection{Characteristics of the Gravitational Wave Parameter Space}
\label{subsec-GW-features}

In the preceding subsections, we have shown that the inverted Type-I 2HDM accommodates a sizable region of parameter space capable of realizing an SFOEWPT.
One of the most compelling phenomenological implications of such a transition is the potential generation of GWs.
In this subsection, we compute the resulting GW spectrum---focusing on its peak amplitude, peak frequency, and the corresponding SNR at LISA---using \package{BSMPT} version~3~\cite{Basler:2024aaf}.
We adopt the default bubble wall velocity $\vwl = 0.95$.
Parameter points with $\text{SNR} > 10$ are identified as GW parameter points, and we analyze their key characteristics to elucidate the structure of the associated GW parameter space.

\begin{figure}[t]
\centering
\includegraphics[width=\textwidth]{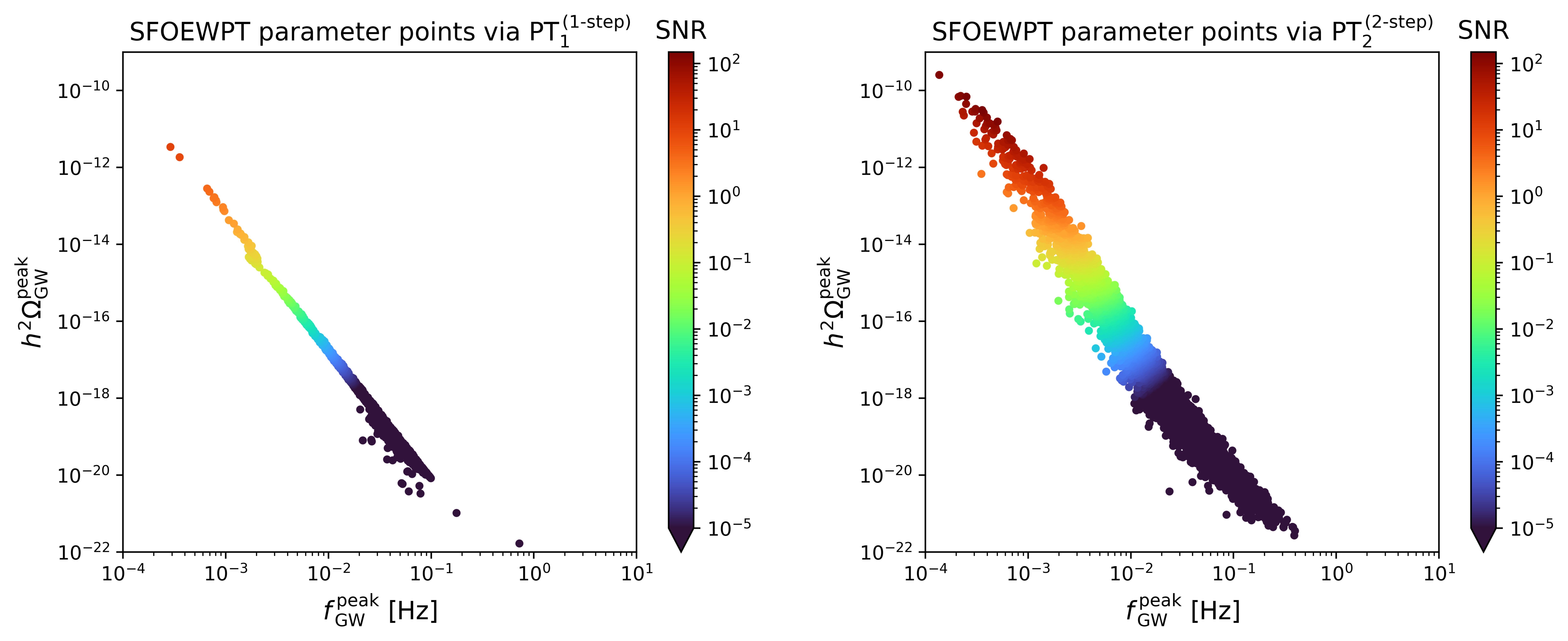} 
\caption{%
	$h^2 \omswp$ versus $\fswp$ for SFOEWPT parameter points, with the color code indicating the SNR (for a four-year LISA mission).
	Results shown are for the dominant sound wave contributions.
	The left and right panels present results for $\trans{1}{1}$ and $\trans{2}{2}$ transitions, respectively.
}
\label{fig-Omegapeak-fpeak-SNR} 
\end{figure}

\autoref{fig-Omegapeak-fpeak-SNR} displays the peak amplitude $h^2 \omswp$ versus the peak frequency $\fswp$ for SFOEWPT parameter points, focusing on the dominant GW contributions from sound waves.
The color scale represents the SNR computed for a four-year LISA mission.
The left panel shows results for the $\trans{1}{1}$ transition, while the right panel corresponds to $\trans{2}{2}$.
A common inverse correlation between $h^2 \omswp$ and $\fswp$ is observed across both transition types: higher peak frequencies generally correspond to lower peak amplitudes.
However, the degree of correlation varies.
The $\trans{1}{1}$ transitions exhibit a tighter correlation, manifested as a narrower distribution, compared to the broader spread seen in the $\trans{2}{2}$ case.

The SNR values at LISA, illustrated by the color map in \autoref{fig-Omegapeak-fpeak-SNR}, span a broad range—from well below $10^{-5}$ to $\mathcal{O}(10^2)$.
The number of GW parameter points satisfying the $\text{SNR} > 10$ criterion varies significantly across different transition types.
Out of the $2.36\times 10^6$ physical parameter points obtained from the random scan over the ranges specified in \autoref{eq-scan-range}, the number of GW parameter points for each transition stage is:
\beq
\label{eq-GW-no}
\begin{alignedat}{3}
N_\text{GW}^{\trans{1}{1} }&= 1,  &\qquad N_\text{GW}^{\trans{1}{2}} &= 0,  \\
N_\text{GW}^{\trans{2}{2}} &= 114, &\qquad 
N_\text{GW}^{\trans{3}{3}} &= 1 .
\end{alignedat}
\eeq
Including the results of the targeted scan for $\trans{3}{3}$—which yielded a total of 29 SFOEWPT parameter points as shown in \autoref{fig-mch-mh-xic}—the number of GW parameter points for this transition slightly increases to $N_\text{GW}^{\trans{3}{3}} = 2$.

This distribution is noteworthy: although one-step transitions account for the largest share of SFOEWPT parameter points (see \autoref{eq-probability}), they rarely produce GW signals detectable by LISA.
In contrast, $\trans{2}{2}$ transitions stand out as the most productive source of potentially observable GW signals in our scan.

\begin{figure}[t]
\centering
\includegraphics[width=\textwidth]{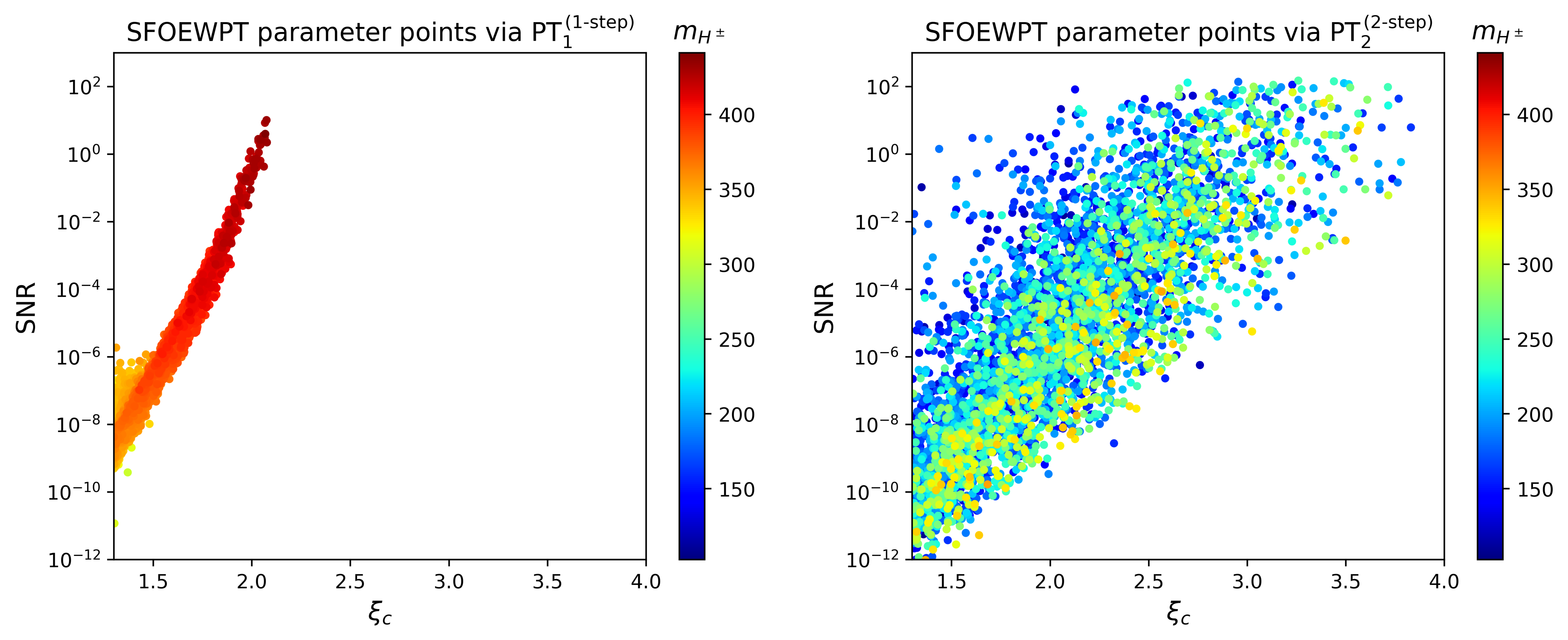} 
\caption{%
	GW SNR at LISA versus $\xi_c$ for SFOEWPT parameter points, with the color code indicating $\mch$ (in units of GeV).
	The left and right panels present results for $\trans{1}{1}$ and $\trans{2}{2}$ transitions, respectively.
}
\label{fig-SNR-xic-mch} 
\end{figure}

We next examine the correlation between the GW SNR at LISA and the phase transition strength $\xi_c$.
A strong correlation between these quantities would be especially valuable: the detection of a high-SNR signal could serve as a robust probe of the underlying electroweak phase transition dynamics.
Such a constraint on $\xi_c$ would, in turn, help to restrict the viable parameter space and improve our ability to discriminate between competing new physics scenarios.
\autoref{fig-SNR-xic-mch} shows the  SNR  versus  $\xi_c$ for SFOEWPT parameter points, with the charged Higgs mass $\mch$ indicated by the color scale.
The left panel corresponds to $\trans{1}{1}$ transitions, while the right panel displays results for $\trans{2}{2}$.

For one-step transitions, a clear positive correlation is observed: larger values of $\xi_c$ generally correspond to higher SNRs.
This suggests that stronger phase transitions are more likely to produce GW signals detectable by LISA.
In contrast, no such strong correlation is evident for $\trans{2}{2}$ transitions.
Even near $\xi_c \approx 2$, the associated SNR values span several orders of magnitude, from $\mathcal{O}(10^{-10})$ to $\mathcal{O}(1)$.
Nonetheless, all GW parameter points with $\text{SNR} > 10$ satisfy $\xi_c > 2$, indicating that a sufficiently strong phase transition is still a necessary condition for GW signal detectability.
The charged Higgs mass $\mch$ shows no significant correlation with SNR in either transition type.
In summary, our results highlight a robust correlation between $\xi_c$ and SNR in one-step transitions, while this relationship appears to be absent or substantially weakened in multi-step scenarios.

\begin{figure}[t]
\centering
\includegraphics[width=\textwidth]{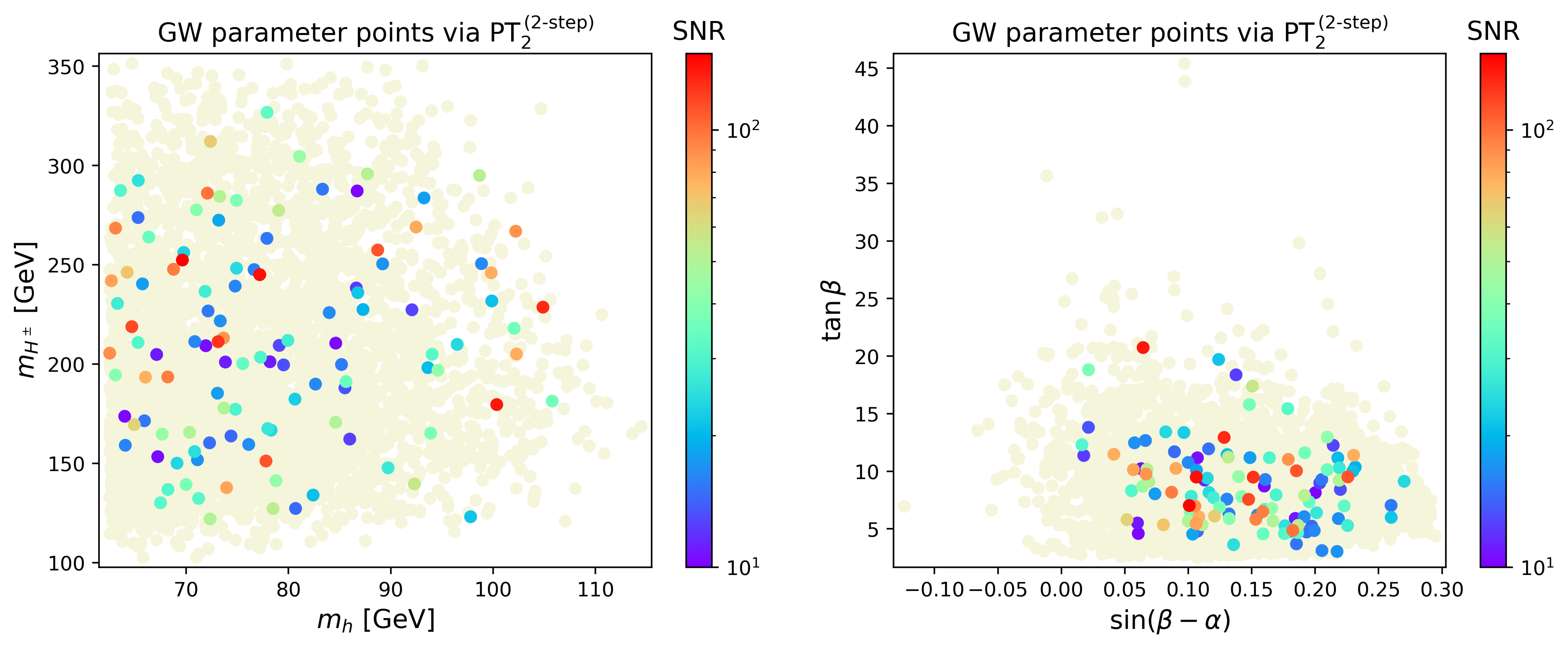} 
\caption{%
	$\mch$ vs $\mh$ for GW parameter points (SFOEWPT points with $\text{SNR}>10$) for $\trans{2}{2}$ transition, with the color code indicating SNR.
	SFOEWPT parameter points that do not satisfy $\text{SNR}>10$ are denoted by beige points.
}
\label{fig-SNR10} 
\end{figure}

Finally, we investigate whether the condition $\text{SNR} > 10$ imposes additional constraints on the parameter space of the inverted Type-I 2HDM.
\autoref{fig-SNR10} displays the distribution of parameter points satisfying $\text{SNR} > 10$ for $\trans{2}{2}$ transitions—the only transition stage yielding a substantial number of such GW parameter points.
SNR values are indicated by the color scale.
The left panel shows the $(\mh,\mch)$ plane, while the right panel presents $(\sba,\tb )$.
For comparison, SFOEWPT parameter points with $\text{SNR} \leq 10$ are shown in beige.

Strikingly, the requirement of $\text{SNR} > 10$ does not significantly alter or further restrict the allowed parameter space.
Rather, the high-SNR points largely overlap with the broader SFOEWPT regions previously identified for the $\trans{2}{2}$ transitions.

\section{Collider Phenomenology at the 1.5 TeV CLIC}
\label{sec-collider}

\subsection{Golden Channels of the SFOEWPT Parameter Space at the 1.5 TeV CLIC}
\label{subsec-golden-channel}

The preceding sections identified parameter regions within the inverted Type-I 2HDM that can induce an SFOEWPT and potentially generate GW signals detectable at LISA.
These regions, which are significantly more constrained than the full physical parameter space, provide compelling targets for experimental investigation.

This section explores the phenomenological signatures associated with these SFOEWPT parameter points at high-energy colliders.\footnote{%
We focus on the SFOEWPT parameter space, as the absence of a detectable GW signal does not diminish the importance of identifying BSM models capable of supporting an SFOEWPT.
}
A primary motivation is that a discovery at a collider would offer crucial supporting evidence, directly linking macroscopic cosmological events such as the EWPT to specific microscopic dynamics.
Conversely, null results from collider searches can help exclude portions of the SFOEWPT parameter space in the inverted Type-I 2HDM.

Among various high-energy colliders, we select the $1.5\TeV$ stage of the CLIC~\cite{Adli:2025swq}.
CLIC is CERN’s proposed linear electron-positron collider, with the $1.5\TeV$ stage designed to extend the physics reach beyond its initial $380\GeV$ phase.
CLIC employs advanced technologies, including a two-beam acceleration scheme in which a separate high-current drive beam generates radio-frequency (RF) power at $12\GHz$ to feed the main accelerating structures. These structures operate with accelerating gradients up to $100\, \text{MV/m}$.
The $1.5\TeV$ configuration features a $29\km$ linac and recombination systems capable of producing 100\,A current pulses.
The design emphasizes modularity, allowing for the reuse of infrastructure from the $380\GeV$ stage and enabling a cost-effective upgrade to higher energies.
The combination of a clean experimental environment and high center-of-mass energy makes the $1.5\TeV$ CLIC stage particularly well-suited for probing the SFOEWPT parameter space of the inverted Type-I 2HDM, where $\mch$ and $\ma$ are bounded from above by approximately $440\GeV$ and $430\GeV$, respectively (see \autoref{eq-summary-param}).

\begin{figure}[t]
\centering
\includegraphics[width=\textwidth]{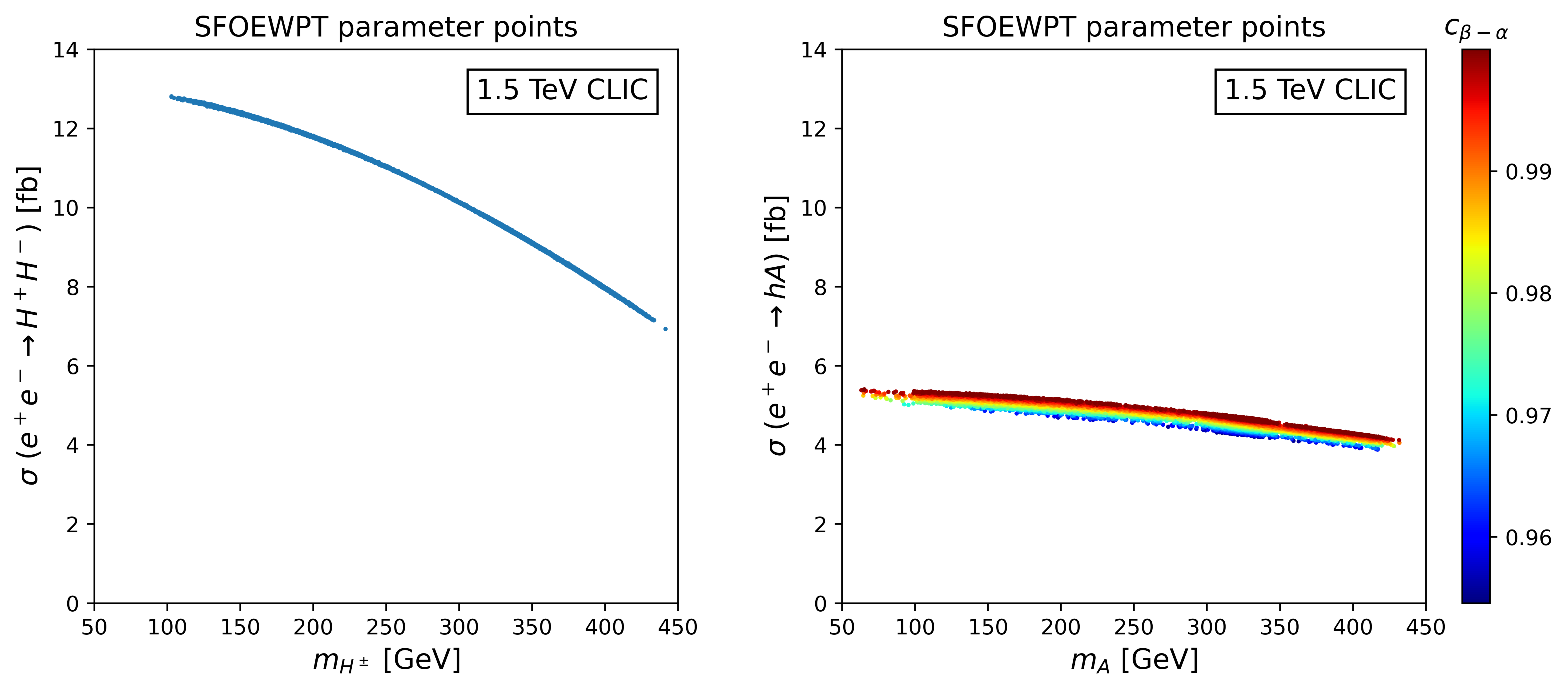} 
\caption{%
	Production cross sections at the $1.5\TeV$ CLIC for $\ee \to H^+ H^-$ (left panel, as a function of $\mch$) and $\ee \to A h$ (right panel, as a function of $\ma$, color-coded by $\cba$) over the SFOEWPT parameter space.
}
\label{fig-xsec-mh}
\end{figure}

Let us first examine the most efficient production channels for BSM Higgs bosons relevant to the SFOEWPT parameter space.
For the lighter \textit{CP}-even Higgs boson $h$, potential production modes include Higgsstrahlung ($\ee \to Z^* \to Z h$), $WW$ fusion ($\ee \to \nu_e \bar{\nu}_e h$), $ZZ$ fusion ($\ee \to e^+ e^- h$), and associated production with top quark pairs ($\ee \to \ttop h$).
However, these processes are typically suppressed in or near the Higgs alignment limit, where the $V$–$V$–$h$ coupling ($V = W, Z$) becomes negligible, or in the large $\tb$ regime, where the Yukawa coupling modifier $\xi^h_f$ is significantly reduced.

Given these limitations, we shift our focus to production channels involving $H^\pm$ and $A$, specifically the pair production process $\ee \to H^+ H^-$ and the associated production $\ee \to A h$.
The cross section for $\ee \to H^+ H^-$, mediated by $\gamma$ and $Z$ exchange, depends predominantly on the charged Higgs mass $\mch$, as detailed in \autoref{eq-gauge-coupling}.
The $\ee \to A h$ channel, proceeding through an $s$-channel $Z$ boson, has a coupling strength governed by the $Z$-$A$-$h$ vertex, which is proportional to $\cba$.

\autoref{fig-xsec-mh} shows the calculated cross sections for the two production modes across the SFOEWPT parameter space.
The left panel displays $\sigma(\ee \to H^+ H^-)$ as a function of $\mch$, while the right panel shows $\sigma(\ee \to A h)$ as a function of $\ma$, with points color-coded by $\cba$.
Both channels can yield sizable production rates at the $1.5\TeV$ CLIC, typically at or above $\mathcal{O}(1)\fb$.
However, the $\ee \to H^+ H^-$ process generally results in significantly larger cross sections and, importantly, depends almost exclusively on $\mch$.
For these reasons, we focus on charged Higgs pair production as the primary channel for probing the SFOEWPT parameter space in the inverted Type-I 2HDM.

\begin{figure}[t]
\centering
\includegraphics[width=0.6\textwidth]{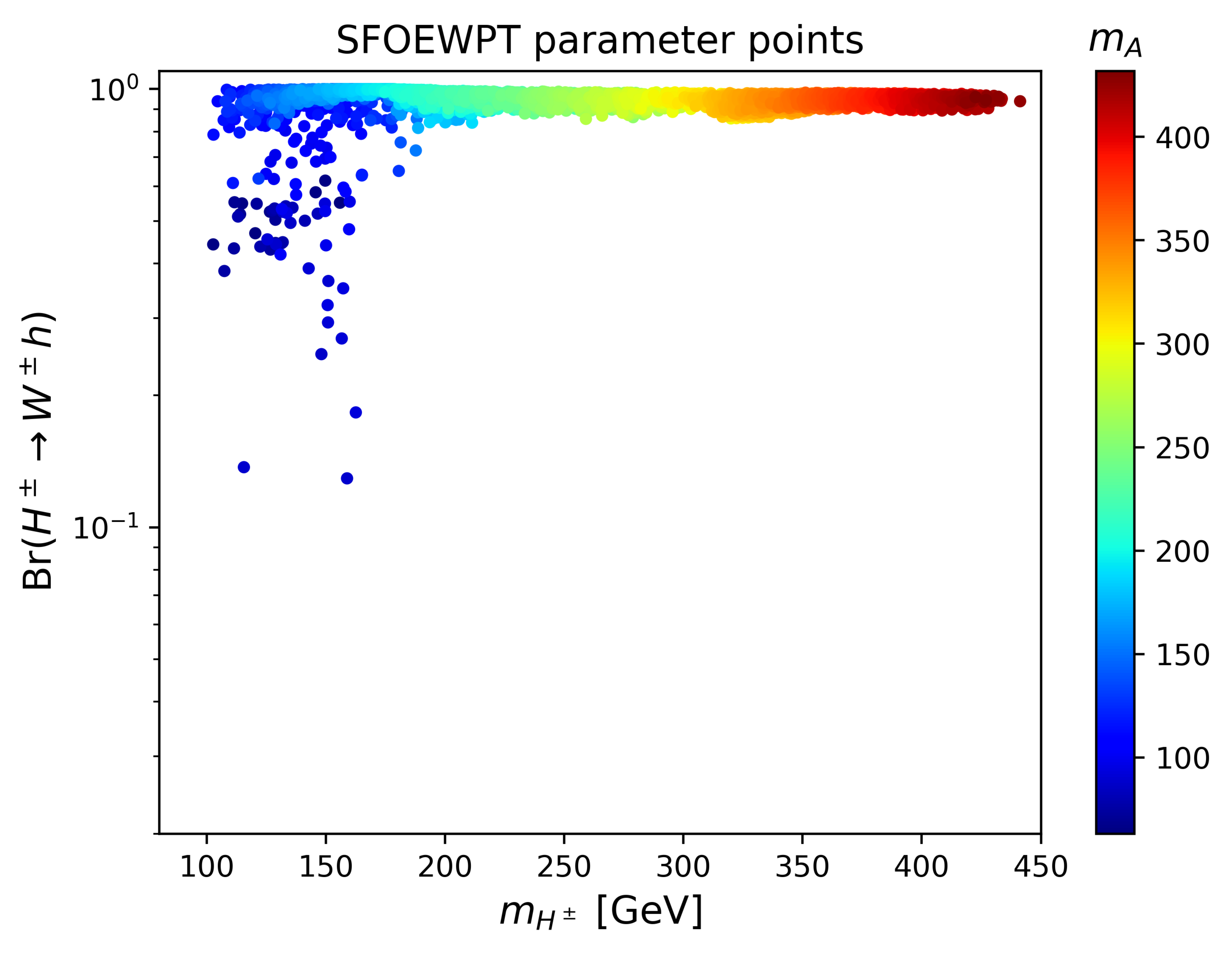} 
\caption{%
	Branching ratio of the decay mode $H^\pm\to W^\pm h$ as a function of $\mch$ across the SFOEWPT parameter space.
	The color code denotes $\ma$ in units of GeV.
}
\label{fig-BRcH-mch-ma}
\end{figure}

We now turn to the decay patterns of $H^\pm$ within the SFOEWPT parameter space.
The charged Higgs boson $H^\pm$ predominantly decays via $H^\pm\to W^\pm h$.
This behavior is illustrated in \autoref{fig-BRcH-mch-ma}, which shows the branching ratio $\br(H^\pm\to W^\pm h)$ as a function of $\mch$, with $\ma$ indicated by the color scale.
For heavier pseudoscalar masses ($\ma \gtrsim 200\GeV$), the $H^\pm\to W^\pm h$ channel is overwhelmingly dominant, with branching ratios typically exceeding 80\%.
For lighter $\ma$, however, a small subset of SFOEWPT parameter points exhibits a reduced $\br(H^\pm\to W^\pm h)$, which can be as low as approximately 10\%; this reduction is primarily due to the competing decay $H^\pm\to W^\pm A$ becoming kinematically accessible.
Nonetheless, for the vast majority of SFOEWPT parameter points, the $H^\pm\to W^\pm h$ mode remains dominant, motivating our focus on the following golden channel for collider studies:
\[
e^+ e^- \to H^+ H^- \to W^+ W^- h h.
\]

The final-state signatures of this golden channel are dictated by the subsequent decays of the lighter \textit{CP}-even Higgs boson $h$, whose branching ratios depend sensitively on the underlying model parameters.
The dominant decay modes typically involve fermion pairs, governed by the Yukawa couplings of $h$, which scale universally with the factor $\xi^h_f$ and are proportional to the corresponding fermion masses  in the Type-I 2HDM.
Additional relevant decay channels include $WW^*$ and $ZZ^*$, which become significant at larger $\mh$---near the respective kinematic thresholds---and for larger values of $|\sba|$, since the $hVV$ $(V = W^\pm, Z)$ coupling scales with $\sba$.

Another potentially significant decay channel for $h$ is $h\to\gamma\gamma$.
The partial decay width is given by~\cite{Djouadi:2005gj}:
\begin{equation}
\label{eq-decay-width-rr}
\Gamma(h\to\gamma\gamma) = \frac{\alpha_\text{em}^2 \mh^3}{256 \pi^3 v^2} 
\Bigl| \sum_f N_C^f Q_f^2 \xi^h_f A^\mathcal{H}_{1/2}(\tau_f) +
	\sba A^\mathcal{H}_1(\tau_W) + \frac{v^2}{2\mch^2} \hat{g}_{h H^+ H^-}  A^\mathcal{H}_0(\tau_{H^\pm})
\Bigr|^2,
\end{equation}
where $N_C^f$ and $Q_f$ are the color factor and electric charge of fermion $f$, respectively; $\xi^h_f$ is defined in \autoref{eq-Yukawa}; and $\tau_i = \mh^2 / (4 m_i^2)$.
The loop functions $A_J^\mathcal{H}(\tau)$ are given by~\cite{Djouadi:2005gj}:
\[
\begin{split}
A^\mathcal{H}_{1/2}(\tau) &= \frac{2}{\tau^2} \bigl[ \tau + (\tau-1) f(\tau) \bigr], \\
A^\mathcal{H}_1(\tau) &= -\frac{1}{\tau^2} \bigl[ 2\tau^2 + 3\tau + 3(2\tau-1) f(\tau) \bigr], \\
A^\mathcal{H}_0(\tau) &=- \frac{1}{\tau^2} \bigl[ \tau-f(\tau) \bigr],
\end{split}
\]
where the function $f(\tau)$ is: 
\[
f(\tau) = \begin{cases} 
	\arcsin^2 \sqrt{\tau} & \text{if $\tau \leq 1$;} \\
	-\frac{1}{4} \Bigl[
		\ln \bigl(\frac{\sqrt{\tau}+\sqrt{\tau-1}}{\sqrt{\tau}-\sqrt{\tau-1}}\bigr) - i\pi
	\Bigr]^2 & \text{if $\tau > 1$.}
\end{cases}
\]
The normalized coupling $\hat{g}_{h H^+ H^-}$ in \autoref{eq-decay-width-rr} is given by~\cite{Bernon:2015qea}:
\[
\hat{g}_{h H^+ H^-} = \frac{1}{v^2}
	\Bigl[ \sba \bigl\{ \mh^2 + 2 (\mch^2-\mbsq) \bigr\} + \cba\, \frac{1-\tb^2}{\tb} (\mh^2-\mbsq) \Bigr].
\]

\begin{figure}[t]
\centering
\includegraphics[width=0.6\textwidth]{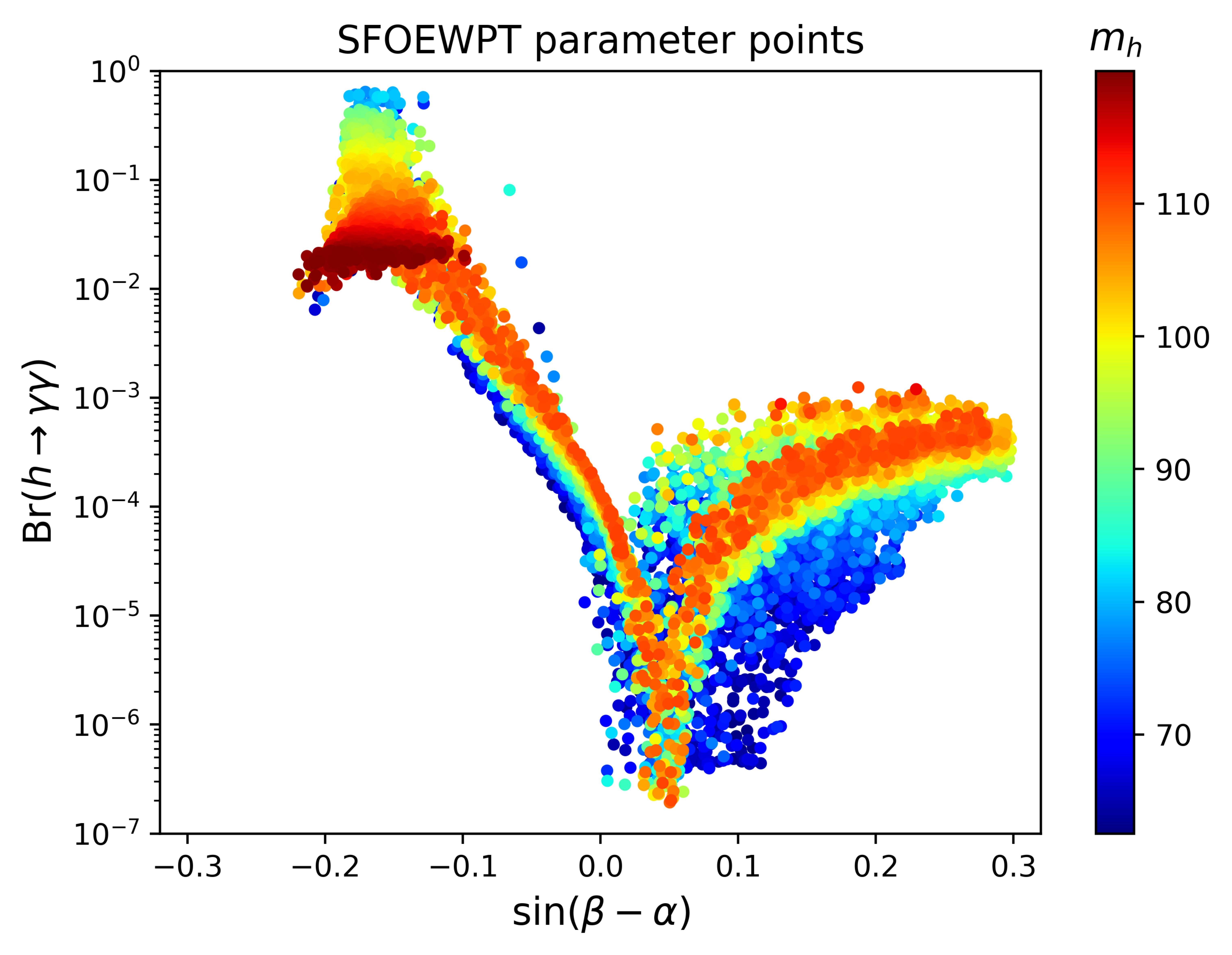} 
\caption{%
	$\br(h\to\gamma\gamma)$ versus $\sba$ for SFOEWPT parameter points.
	The color code denotes $\mh$ in units of GeV.
	Parameter points are sorted by $\mh$, with smaller $\mh$ values plotted underneath.
}
\label{fig-Brr-sba-mh-PT11}
\end{figure}

Interestingly, $\br(h\to\gamma\gamma)$ exhibits a strong sensitivity to $\sba$, particularly its sign.
\autoref{fig-Brr-sba-mh-PT11} displays $\br(h\to\gamma\gamma)$ as a function of $\sba$ across the SFOEWPT parameter points, with $\mh$ indicated by the color scale (points with smaller $\mh$ are plotted beneath those with larger $\mh$).
These branching ratios are computed using \package{2HDMC}~\cite{Eriksson:2009ws}.

For positive $\sba$, the $h\to\gamma\gamma$ decay mode is generally negligible, with branching ratios typically below approximately $10^{-3}$.
In contrast, for negative $\sba$, $\br(h\to\gamma\gamma)$ can be significantly enhanced, reaching values as high as $\sim 80\%$.
This enhancement for negative $\sba$ arises primarily from the suppression of $h \to b\bar{b}$ decays.
Since the Yukawa coupling modifier is given by $\xi^h_f = \sba + \cba/\tb$ (see \autoref{eq-Yukawa}), a negative $\sba$ can induce a cancellation with the $\cba/\tb$ term, leading to a suppressed $\xi^h_f$ and hence a reduced $\Gamma(h \to f\bar{f})$.
This suppression, in turn, boosts the relative importance of $h\to\gamma\gamma$.
Given that only one-step SFOEWPTs accommodate negative $\sba$ values (see \autoref{fig-tb-sba-xic}), an observed enhancement in $\br(h\to\gamma\gamma)$ could serve as an indirect probe favoring one-step SFOEWPT realizations.

\begin{figure}[t]
\centering
\includegraphics[width=\textwidth]{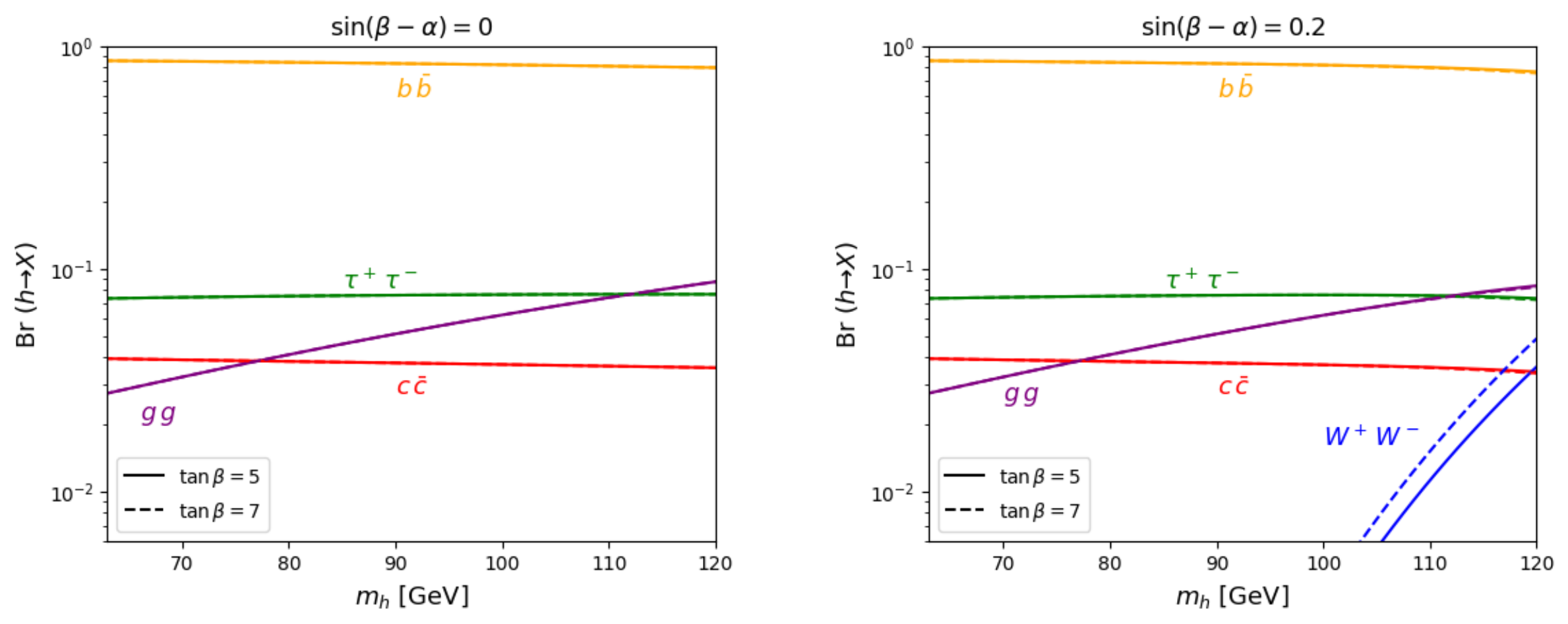} 
\caption{%
	Branching ratios of various decay modes of the lighter \textit{CP}-even Higgs boson $h$ as a function of $\mh$ for $\sba=0$ (left panel) and $\sba=0.2$ (right panel).
	Results are shown for $\tb=5$ (solid lines) and $\tb=7$ (dashed lines), with $\ma=\mch=300\GeV$ and $\mb = \mh$.
}
\label{fig-BR-h-positive-sba}
\end{figure}

We now examine the complete decay pattern of the lighter \textit{CP}-even Higgs boson $h$.
First, \autoref{fig-BR-h-positive-sba} shows two cases: the exact alignment limit with $\sba = 0$ (left panel) and a moderate positive deviation with $\sba = 0.2$ (right panel).
The branching ratios of $h$ are plotted as a function of $\mh$.
We consider $\tb = 5$ (solid lines) and $\tb = 7$ (dashed lines), setting $\ma = \mch = 300\GeV$ and $\mb = \mh$.
The condition $\mb = \mh$ is imposed to avoid theoretical restrictions on $\tb$~\cite{Gu:2017ckc,Dorsch:2017nza,Su:2020pjw}.

For both $\sba = 0$ and $\sba = 0.2$, the dominant decay mode of $h$ is into $b\bar{b}$, followed by $\tau^+\tau^-$.
The third most significant decay mode is $c\bar{c}$ for $\mh \lesssim 76\GeV$, and $gg$ for $76\GeV \lesssim \mh \lesssim 110\GeV$.
The fermionic branching ratios show little dependence on $\tb$, as evidenced by the close overlap of solid and dashed lines.
The $h \to WW^*$ channel is absent in the exact Higgs alignment limit ($\sba = 0$) but becomes sizable when $\mh \gtrsim 105\GeV$ and $\sba=0.2$.
At larger $\tb$, the suppression of fermionic decays (due to the $\cba/\tb$ term in $\xi_h^f$) enhances the relative importance of $h \to WW^*$.

\begin{figure}[t]
\centering
\includegraphics[width=\textwidth]{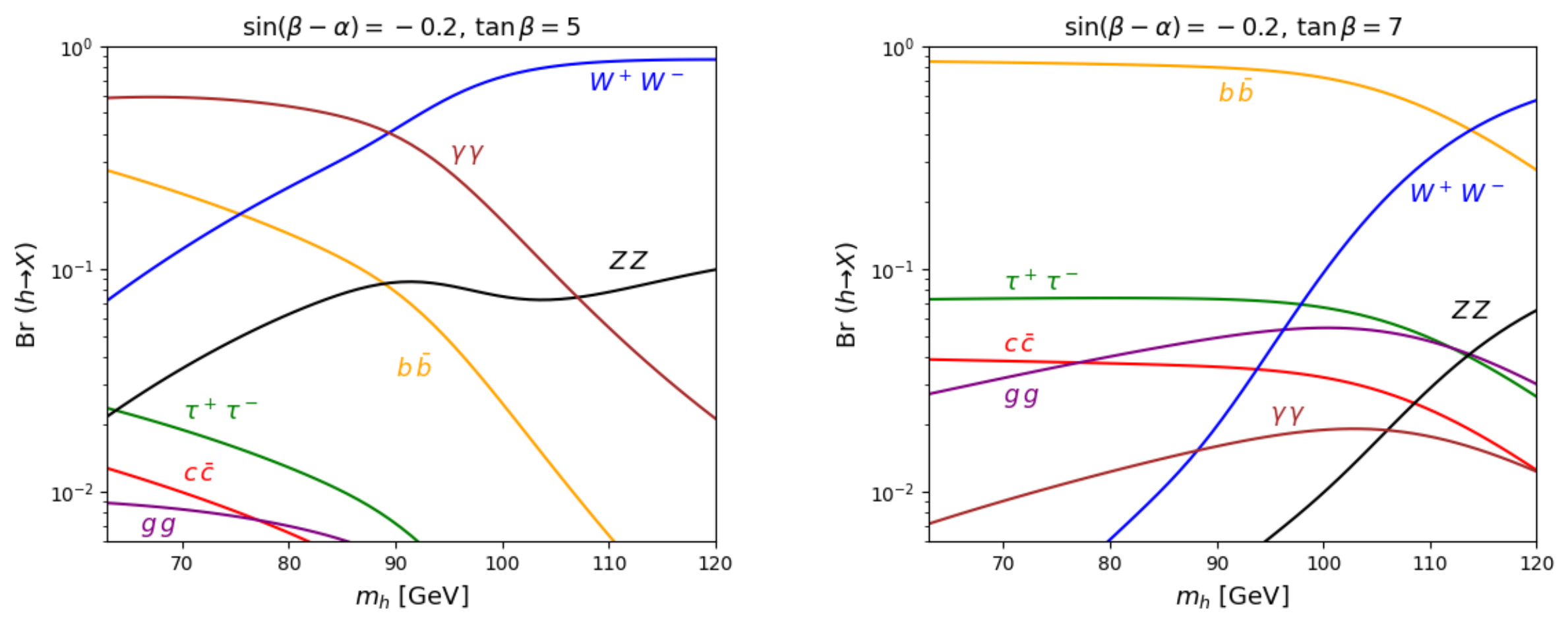} 
\caption{%
	For negative $\sba=-0.2$, branching ratios of various decay modes of the lighter \textit{CP}-even Higgs boson $h$ as a function of $\mh$.
	Results are shown for $\tb=5$ (left panel) and $\tb=7$ (right panel), with $\ma=\mch=300\GeV$ and $\mb = \mh$.
}
\label{fig-BR-h-negative-sba}
\end{figure}

For negative $\sba$, the decay pattern of $h$ changes considerably compared to the positive or aligned cases.
\autoref{fig-BR-h-negative-sba} presents the branching ratios for $\sba = -0.2$.
Here, results are shown for $\tb = 5$ (left panel) and $\tb = 7$ (right panel), assuming $\ma = \mch = 300\GeV$ and $\mb = \mh$.
For $\tb = 5$ and $\mh \lesssim 90\GeV$, $h\to\gamma\gamma$ can become the dominant decay mode, followed by $h \to b\bar{b}$, $h \to \tau^+\tau^-$, $h \to c\bar{c}$, and $h \to gg$.
This dramatic enhancement of $h\to\gamma\gamma$ arises from a near cancellation in the Yukawa coupling modifier $\xi^h_f$ (leading to $\xi^h_f \approx -0.004$ for $\sba = -0.2$ and $\tb = 5$, for example), which drastically suppresses fermionic decays.
For heavier $h$ in the range $90\GeV \lesssim \mh \lesssim 120\GeV$, the dominant decay mode shifts to $WW^*$, with $h\to\gamma\gamma$ and $h\to ZZ^*$ also contributing significantly.

For $\tb = 7$ (right panel of \autoref{fig-BR-h-negative-sba}), $h \to b\bar{b}$ generally remains dominant for $\mh \lesssim 90\GeV$, followed by $\tau^+\tau^-$ and $c\bar{c}/gg$.
Although $h\to\gamma\gamma$ is typically subdominant in this case, its branching ratio can still reach a few percent.
When $\mh \gtrsim 90\GeV$, $WW^*$ and $ZZ^*$ modes become important, while $\br(h\to\gamma\gamma)$ can persist at the $\mathcal{O}(10^{-2})$ level.
In summary, the $h\to\gamma\gamma$ decay mode attains a significantly enhanced branching ratio---potentially reaching several percent or more---if $\sba < 0$, providing a distinctive feature of such scenarios.

Considering the dominant $H^\pm\to W^\pm h$ decay and the characteristic decay patterns of $h$ (which depend on the sign of $\sba$), we propose the following two final states  to probe the SFOEWPT parameter space of the inverted Type-I 2HDM at CLIC:
\begin{align}
\label{eq-channel-bbttau}
e^+ e^- &\to H^+ H^- \to W^+ W^- h h \to W^+ W^- b\bar{b}\tau^+\tau^-, \\
\label{eq-channel-bbrr} 
e^+ e^- &\to H^+ H^- \to W^+ W^- h h \to W^+ W^- b\bar{b}\gamma\gamma.
\end{align}
The process in \autoref{eq-channel-bbrr} is particularly interesting for targeting the negative $\sba$ case, which predominantly corresponds to one-step SFOEWPT scenarios.

\subsection{Signal-to-Background Analysis}
\label{subsec-signal-analysis}

In this subsection, we perform a detector-level signal-to-background analysis for the proposed golden channels defined in Eqs.~\eqref{eq-channel-bbttau} and \eqref{eq-channel-bbrr}.
A key aspect of evaluating the discovery potential of the $1.5\TeV$ CLIC is the precise estimation of SM backgrounds for the complex final states $W^+ W^- b\bar{b}\tau^+\tau^-$ and $W^+ W^- b\bar{b} \gamma \gamma$.

However, direct computation of the SM cross sections for these final states using public Monte Carlo tools—such as \package{MadGraph5\_aMC@NLO}\cite{Alwall:2011uj} or \package{CalcHEP}\cite{Belyaev:2012qa}—is often practically infeasible.
These simulations can fail to converge even after extensive runtimes (exceeding ten days), primarily due to the enormous number of Feynman diagrams contributing to such high-multiplicity processes.
For example, the $W^+ W^- b\bar{b}\gamma\gamma$ final state involves 4,308 tree-level diagrams, which significantly complicates event generation.

To estimate the SM background contributions, we instead compute parton-level cross sections for simplified reference processes with fewer final-state particles: $\ee \to \ww b\bar{b}$ and $\ee \to \ww b\bar{b}\gamma$.
Applying representative kinematic cuts (for $b$-quarks: $p_T^b > 20 \GeV$, $|\eta_b| < 2.5$, and $m_{b\bar{b}} < \mh$;\footnote{%
	The cut $m_{b\bar{b}} < \mh$ is primarily intended to enhance signal characteristics and will be further justified in the subsequent signal selection analysis.
}
for photons in the second process: $p_T^\gamma > 10\GeV$ and $|\eta_\gamma| < 2.5$), we obtain the following cross sections:
\beq
\label{eq-sigma-SM}
\begin{split}
\sigma_\text{SM}(\ee \to \ww b\bar{b})\big|_\text{cuts} &\simeq 6.79\fb, \\ 
\sigma_\text{SM}(\ee \to \ww b\bar{b}\gamma)\big|_\text{cuts} &\simeq 0.782\fb.
\end{split}
\eeq

We expect the full SM background cross section for $e^+ e^- \to W^+ W^- b\bar{b}\tau^+\tau^-$ to be substantially smaller than $\sigma_{\rm SM}(\ee \to \ww b\bar{b})$.
This suppression arises from additional electroweak coupling factors (e.g., proportional to $\alpha_{\text{em}}^2$ for the $\tau^+\tau^-$ pair) and further phase-space suppression due to the increased final-state multiplicity.
An analogous suppression is expected when comparing $e^+ e^- \to W^+ W^- b\bar{b}\gamma\gamma$ to the simpler process $e^+ e^- \to W^+ W^- b\bar{b}\gamma$.

It is therefore reasonable to expect the cross sections for the full SM background processes $e^+ e^- \to W^+ W^- b\bar{b}\tau^+\tau^-$ and $e^+ e^- \to W^+ W^- b\bar{b}\gamma\gamma$ to lie well below $1\ab$.
Given the anticipated total integrated luminosity of $4\iab$ at the $1.5\TeV$ CLIC, the expected number of background events is significantly less than one, rendering these backgrounds effectively negligible for our signal analysis.

A key question then arises: how should the signal significance be computed when the expected number of background events ($N_\text{bg}$) is negligible?
Consider the general case in which $N_\text{tot}$ events are observed, given an SM background prediction for $N_\text{bg}$ (assuming, for simplicity, no uncertainty on $N_\text{bg}$).
Since $N_\text{tot} = N_\text{sg} + N_\text{bg}$, where $N_\text{sg}$ denotes the number of signal events, the significance $\mathcal{S}$ based on a Poisson counting likelihood is given by~\cite{ATLAS:2020yaz}:
\begin{equation}
\label{eq-significance-general}
\mathcal{S} = \sqrt{2 \Bigl\{ N_\text{tot} \ln \frac{N_\text{tot}}{N_\text{bg}} - N_\text{sg} \Bigr\} }.
\end{equation}
In scenarios with a sufficiently large $N_\text{bg}$ (typically $\gtrsim 20$), where the Gaussian approximation to the Poisson distribution is valid, this expression simplifies to:
\[
\mathcal{S}\big|_\text{Gaussian} \approx \frac{N_\text{sg}}{\sqrt{N_\text{bg}}}.
\]

However, when $N_\text{bg}$ is small, the Gaussian approximation becomes unreliable.
This is because it incorrectly assumes symmetric fluctuations in the background, whereas the Poisson distribution that governs background-only expectations is inherently asymmetric at low event counts---especially in the tails, which are critical for $p$-value estimation.
In such cases, the Poisson-based likelihood formula in \autoref{eq-significance-general} is theoretically more accurate.

Then, a complication arises when $N_\text{bg}$ approaches or equals zero.
Specifically, if $N_\text{bg} \to 0$, the logarithmic term in \autoref{eq-significance-general} causes the significance $\mathcal{S}$ to diverge.
The expression becomes ill-defined for $N_\text{bg} = 0$.
Given these considerations---and in light of our earlier estimate indicating negligible background contributions for the proposed golden channels---we choose to present only the expected number of signal events ($N_\text{sg}$), rather than quoting significance values that would be ill-defined or potentially misleading in the $N_\text{bg} \approx 0$ regime.

\begin{table}[t]
\centering
\setlength{\tabcolsep}{5pt}
\setlength{\heavyrulewidth}{.08em}
\setlength{\lightrulewidth}{.05em}
\renewcommand{\arraystretch}{1.3}
\begin{tabular}{|c||r|r|r|r|r|r||c|c||c|c|} 
\toprule
BP & \multicolumn{1}{c|}{$\mch$} & \multicolumn{1}{c|}{$\mh$} &
    \multicolumn{1}{c|}{$\ma$} & \multicolumn{1}{c|}{$\sba$} &
    \multicolumn{1}{c|}{$\tb$} & \multicolumn{1}{c||}{$\mb$} &
    $\br(h\to bb)$ & $\br(h\to \gamma\gamma)$ & Stage & $\xi_c$ \\
\midrule
BP-1 & 104.6 & 71.3 & 107.5 & 0.022 & 11.8 & 72.0 & $8.42\times 10^{-1}$ & $3.00\times 10^{-6}$ & $\trans{2}{2}$  & 2.23 \\
BP-2 & 199.1 & 70.4 & 168.7 & 0.151 & 4.5 & 68.6 & $8.42\times 10^{-1}$ & $3.50\times 10^{-5}$ & $\trans{2}{2}$  & 1.31 \\
BP-3 & 199.4 & 101.2 & 174.3 & 0.184 & 12.6 & 95.9 & $8.02\times 10^{-1}$ & $6.39\times 10^{-4}$ & $\trans{2}{2}$  & 1.69 \\
BP-4 & 301.3 & 71.2 & 300.7 & 0.037 & 8.6 & 51.2 & $8.42\times 10^{-1}$ & $3.00\times 10^{-6}$ & $\trans{2}{2}$ & 1.47 \\
BP-5 & 309.2 & 97.7 & 313.5 & 0.217 & 9.9 & 90.4 & $8.09\times 10^{-1}$ & $5.07\times 10^{-4}$ & $\trans{2}{2}$  & 2.80 \\      
BP-6 & 345.5 & 78.0 & 343.8 & $-0.171$ & 5.7 & 52.6 & $3.52\times 10^{-2}$ & $6.42\times 10^{-1}$ & $\trans{1}{1}$  & 1.41 \\
BP-7 & 351.1 & 79.8 & 338.7 & 0.200 & 6.8 & 61.4 & $8.33\times 10^{-1}$ & $1.82\times 10^{-4}$ & $\trans{2}{2}$ & 2.57 \\
BP-8 & 395.1 & 102.7 & 402.5 & 0.004 & 5.8 & 85.6 & $8.06\times 10^{-1}$ & $9.10\times 10^{-5}$ & $\trans{1}{1}$  &  1.77 \\
BP-9 & 401.8 & 95.6 & 399.2 & $-0.186$ & 5.6 & 70.6 & $1.90\times 10^{-1}$ & $2.02\times 10^{-1}$ & $\trans{1}{1}$ & 1.77 \\
BP-10 & 403.1 & 119.8 & 393.6 & $-0.178$ & 4.6 & 102.6 & $1.53\times 10^{-1}$ & $1.86\times 10^{-2}$ & $\trans{1}{1}$  & 1.77 \\   \bottomrule
\end{tabular}
\caption{%
	Representative benchmark points (BPs) from the SFOEWPT parameter space, illustrating model parameters, key branching ratios (BRs) of the lighter \textit{CP}-even Higgs boson $h$, the associated transition stage, and the corresponding value of $\xi_c$.
	All masses ($\mch$, $\mh$, $\ma$, $\mb$) are given in GeV.
} 
\label{tab-benchmarks}
\end{table}

To efficiently explore the SFOEWPT parameter space in our phenomenological study, we select ten representative benchmark points (BPs), summarized in \autoref{tab-benchmarks}.
The table lists the model parameters, the key branching ratios of the lighter \textit{CP}-even Higgs boson $h$, the stage of electroweak phase transition, and the corresponding value of $\xi_c$.
Among these, BP-6, BP-8, BP-9, and BP-10 realize an SFOEWPT via a one-step transition, while the remaining points correspond to $\trans{2}{2}$.
Notably, BP-6, BP-9, and BP-10 feature negative values of $\sba$.

For these benchmark points, we conduct a detector-level signal analysis.
Event generation is performed using \package{MadGraph5\_aMC@NLO}.
Parton showering and hadronization are handled by \package{Pythia} version 8.309~\cite{Bierlich:2022pfr}, followed by fast detector simulation using \package{Delphes} version 3.5.0~\cite{deFavereau:2013fsa} with the \filename{delphes\_card\_CLICdet\_Stage2.tcl} configuration.
Jet clustering is carried out using the exclusive Valencia\footnote{%
	The Valencia algorithm is particularly well-suited for high-energy lepton colliders, as it effectively mitigates the impact of initial-state radiation and beam-induced backgrounds by reconstructing beam jets explicitly.
	We adopt the exclusive mode, which halts jet clustering once the desired number of jets is reached—appropriate here given the known jet multiplicity.
}
algorithm~\cite{Boronat:2014hva,Boronat:2016tgd}, as implemented in \package{FastJet}~\cite{Cacciari:2011ma}.
The algorithm is configured with standard parameters: jet radius $R = 0.5$, and energy/angular weighting parameters $\beta = \gamma = 1$, which define the distance metric used in clustering~\cite{Boronat:2016tgd}.

Let us first present the signal analysis for the $hh\to b\bar{b}\tau^+\tau^-$ mode, focusing on the final state:
\[
e^+ e^- \to H^+ H^- \to W^+ W^- h h \to \ell^\pm \nu j j b\bar{b}\tau^+\tau^-,
\]
where $\ell^\pm = e^\pm,\mu^\pm$.
This final state comprises one charged lepton, missing energy attributed to the neutrino (reconstructable due to the clean initial state at CLIC), two light-flavor jets, two $b$-tagged jets, and a pair of tau leptons.
For convenience, we refer to this as the $bb\tau\tau$ mode in the discussion that follows.

Accurate reconstruction of this final state critically depends on efficient $b$-tagging and $\tau$-tagging.
A jet is identified as a $b$-jet if a $B$-hadron with $p_T > 5\GeV$ is found within a cone of $\Delta R = 0.3$ around the jet axis.
We implement $b$-tagging efficiencies and mistagging rates based on the default configuration in \filename{delphes\_card\_CLICdet\_Stage2.tcl}, which offers three fixed values for the $b$-tagging efficiency: $P_{b \rightarrow b} = 50\%, 70\%, 90\%$.
For each chosen $P_{b \rightarrow b}$, the corresponding mistagging rates---dependent on jet kinematics such as energy and pseudorapidity---are automatically determined.
In our analysis, we adopt $P_{b \rightarrow b} = 70\%$.
Tau leptons are identified through their hadronic decays, denoted as $\tauh$, which appear as narrow jets with a small number of charged tracks~\cite{CMS:2007sch,Bagliesi:2007qx,CMS:2018jrd}.
Tau-tagging and mistagging rates are also $p_T$-dependent, and we follow the default parametrization provided in \filename{delphes\_card\_CLICdet\_Stage2.tcl} for their implementation.

For the $bb\tau\tau$ mode, we establish the following selection criteria:
\begin{itemize}
  \item \textbf{Lepton Multiplicity:} Require exactly one charged lepton (electron or muon) with transverse momentum $p_T > 10\GeV$ and pseudorapidity $|\eta| < 2.5$.
  \item \textbf{Jet Multiplicity:} Using the \enquote{exclusive} Valencia jet algorithm with the total jet number fixed to six, we require exactly two $b$-tagged jets, two $\tauh$-tagged jets, and two light-flavor jets, each with $p_T > 20\GeV$ and $|\eta| < 2.5$.
   \item \textbf{Invariant Mass Cuts:} Impose upper bounds on the invariant masses of the $b$-jet pair and the $\tauh$-jet pair: $m_{b\bar{b}} < \mh$ and $m_{\tauh\tauh} < \mh$.
\end{itemize}
We opt for upper-only cuts on $m_{b\bar{b}}$ and $m_{\tauh\tauh}$ rather than symmetric mass windows centered around $\mh$.
Our analysis of the reconstructed invariant mass distributions reveals that both $m_{b\bar{b}}$ and $m_{\tauh\tauh}$ exhibit significant skewness toward lower values, rather than peaking sharply at $\mh$.
For $b$-jets, this distortion arises from neutrinos produced in semi-leptonic $B$-hadron decays, finite jet energy resolution, and final-state radiation escaping the jet cone~\cite{Proissl:2014lcz,Abramowicz:2016zbo}.
Similarly, the $m_{\tauh\tauh}$ distribution is affected by neutrinos in all $\tau$ decays, the narrow and low-multiplicity nature of $\tauh$ jets, calorimeter resolution effects, and final-state radiation.

For the $hh \to b\bar{b}\gamma\gamma$ mode, we consider the process:
\[
e^+ e^- \to H^+ H^- \to W^+ W^- h h \to \ell^\pm \nu j j b\bar{b}\gamma\gamma.
\]
The final state consists of one charged lepton, missing energy, two light-flavor jets, two $b$-tagged jets, and a pair of photons.
As in the $bb\tau\tau$ case, we refer to this final state as the $bb\gamma\gamma$ mode.

The selection criteria for the $bb\gamma\gamma$ mode are as follows:
\begin{itemize}
\item \textbf{Lepton Multiplicity:} Require exactly one charged lepton (electron or muon) with $p_T > 10\GeV$ and $|\eta| < 2.5$.
\item \textbf{Jet Multiplicity:} Using the \enquote{exclusive} Valencia jet algorithm with the total number of jets fixed to four, require exactly two $b$-tagged jets and two light-flavor jets, each with $p_T > 20\GeV$ and $|\eta| < 2.5$.
\item \textbf{Photon Multiplicity:} Require exactly two photons with $p_T > 10\GeV$ and $|\eta| < 2.5$.
\item \textbf{Invariant Mass Cuts:} Impose upper bounds on the invariant masses: $m_{b\bar{b}} < \mh$ and $m_{\gamma\gamma} < \mh$.
\end{itemize}

\begin{table}[t]
\centering
\setlength{\tabcolsep}{5pt}
\setlength{\heavyrulewidth}{.08em}
\setlength{\lightrulewidth}{.05em}
\begin{tabular}{|c|r|r||c|r|r|}
\toprule
\multicolumn{6}{|c|}{Number of signal events for $e^+ e^- \to H^+ H^-$} \\
\multicolumn{6}{|c|}{at the $1.5\TeV$ CLIC with $\mathcal{L}_\text{tot}=4\iab$} \\
\midrule
Benchmark & \multicolumn{1}{c|}{$bb\tau\tau$ mode} & \multicolumn{1}{c||}{$bb\gamma\gamma$ mode} &
Benchmark & \multicolumn{1}{c|}{$bb\tau\tau$ mode} & \multicolumn{1}{c|}{$bb\gamma\gamma$ mode} \\
\midrule
BP-1  & 68.49 & 0.00    &  BP-6  & 0.06  & 36.14 \\
BP-2  & 46.64 & 0.05    &  BP-7  & 29.64 & 0.26  \\
BP-3  & 68.50 & 0.98    &  BP-8  & 34.10 & 0.12  \\
BP-4  & 38.55 & 0.00    &  BP-9  & 1.53  & 54.16 \\
BP-5  & 42.38 & 0.75    &  BP-10 & 1.17  & 4.35 \\
\bottomrule
\end{tabular}
\caption{%
	Number of expected signal events for the process $e^+ e^- \to H^+ H^-$ leading to the $bb\tau\tau$ and $bb\gamma\gamma$ final states (defined in text) at the $1.5\TeV$ CLIC with an integrated luminosity of $4\iab$.
	These counts are for the benchmark points detailed in \autoref{tab-benchmarks} and after applying the selection criteria.
}
\label{tab-sig-numbers}
\end{table}

\autoref{tab-sig-numbers} presents the expected number of signal events at the $1.5\TeV$ CLIC with a total luminosity of $4\iab$ for the ten benchmark points in the two golden channels, after all selection cuts.
These counts directly reflect the observational potential at CLIC, given our earlier arguments that the SM backgrounds are expected to be negligible.

Several benchmark points show significant yields in the $bb\tau\tau$ mode.
Notably, BP-1 and BP-3 predict the highest yields with approximately 68.5 events each.
Other points, such as BP-2 (46.6 events), BP-5 (42.4 events), BP-4 (38.6 events), BP-8 (34.1 events), and BP-7 (29.6 events), also yield tens of events, indicating good prospects for observation.
Conversely, BP-6, BP-9, and BP-10 show very few events in this channel ($<2$ events), making their detection challenging via this mode.

The $bb\gamma\gamma$ mode exhibits a complementary pattern to the $bb\tau\tau$ channel.
The benchmark points that yield strong signals in the $bb\tau\tau$ mode---such as BP-1, BP-2, BP-3, and BP-4---produce very few or no events in the $bb\gamma\gamma$ channel.
Conversely, BP-9 (54.2 events) and BP-6 (36.1 events), which were weak in the $bb\tau\tau$ mode, predict substantial event yields in the $bb\gamma\gamma$ channel.
This stark contrast highlights the importance of analyzing both final states to ensure broader coverage of the SFOEWPT parameter space.

From the perspective of observational prospects, most benchmark points yield a substantial number of signal events---typically tens---in at least one of the two proposed golden channels.
Specifically, BP-1, BP-2, BP-3, BP-4, BP-5, BP-7, and BP-8 show strong potential in the $bb\tau\tau$ mode, while BP-6 and BP-9 are particularly promising in the $bb\gamma\gamma$ mode.
Even BP-10, which yields the fewest events overall, predicts a combined total of approximately 5.5 signal events across both channels.
Assuming negligible background,  these signal yields are generally sufficient to indicate strong discovery potential and enable detailed phenomenological studies at the $1.5\TeV$ CLIC.

In conclusion, the projected signal event counts demonstrate that the $1.5\TeV$ CLIC, with an integrated luminosity of $4\iab$, possesses significant discovery potential for the SFOEWPT parameter space within the inverted Type-I 2HDM.
While the $bb\tau\tau$ mode appears broadly promising for a number of benchmark points, the $bb\gamma\gamma$ channel provides crucial complementarity, essential for enhancing the overall discovery potential.
Furthermore, the varying signal strengths and channel preferences across different benchmark points suggest that observations (or non-observations) in these channels could play a vital role in characterizing and potentially differentiating between various SFOEWPT scenarios within the model.

\section{Conclusion} 
\label{sec-conclusion}

In this work, we have performed a comprehensive investigation of the electroweak phase transition (EWPT) within the inverted Type-I two-Higgs-doublet model (2HDM), in which the observed $125\GeV$ Higgs boson corresponds to the heavier \textit{CP}-even scalar $H$.
Motivated by the critical role of a strong first-order EWPT (SFOEWPT) in enabling electroweak baryogenesis, as well as by its associated gravitational wave (GW), we systematically explored the viable parameter space under current theoretical and experimental constraints.
We provided a detailed characterization of multi-step EWPT dynamics, evaluated GW signals, and assessed collider discovery prospects.

Our analysis offers several novel insights into the EWPT within the inverted Type-I 2HDM.
First, we have systematically mapped—for the first time—the parameter space compatible with SFOEWPTs, explicitly including multi-step transitions.
Using the notation $\trans{i}{n}$ to denote the $i$-th transition in an $n$-step sequence, we demonstrated that SFOEWPTs can occur not only via conventional one-step pathways, but also prominently through $\trans{2}{2}$, and occasionally via $\trans{3}{3}$.

A key finding is that parameter regions supporting one-step and two-step SFOEWPTs display distinct characteristics, despite limited overlap.
For instance, the charged Higgs mass $\mch$ in one-step scenarios is tightly constrained to approximately $[295, 441]\GeV$, whereas $\trans{2}{2}$ transitions allow a broader and lighter range, $\mch \in [100, 350]\GeV$.
Similarly, one-step transitions prefer a narrower range of $\tan\beta$ ($\tb \in [4.2, 8.8]$), while $\trans{2}{2}$ transitions accommodate a significantly wider range ($\tb \in [2.5, 45.4]$).

We also presented the first detailed calculation of the GW signal-to-noise ratio (SNR) at LISA for SFOEWPTs in this model, explicitly addressing multi-step transitions.
Although one-step SFOEWPTs are more abundant across the parameter space, points generating detectable GW signals ($\text{SNR}>10$) arise predominantly from $\trans{2}{2}$ transitions.
This underscores the importance of considering multi-step dynamics when assessing GW detectability in extended scalar sectors.

Additionally, we analyzed the correlation between the vacuum uplifting measure $\dfz$ and the EWPT strength parameter $\xi_c$.
We confirmed the clear positive correlation for one-step transitions in the inverted scenario,
which was established  for one-step transitions in the normal scenario.
Crucially, however, we found that this correlation breaks down entirely for multi-step transitions, demonstrating that $\dfz$ cannot serve as a universal proxy for the phase transition strength.
A full finite-temperature analysis thus remains essential whenever intermediate metastable vacua appear.

Finally, we conducted the first collider phenomenology study targeting parameter points compatible with SFOEWPTs in the inverted Type-I 2HDM, focusing on their discovery potential at the proposed $1.5\TeV$ CLIC.
We identified charged Higgs pair production, $e^+e^- \to H^+H^-$, followed by $H^\pm \to W^\pm h$, as a particularly promising discovery channel due to its sizable cross section and no dependence on scalar mixing parameters at tree level.

An important phenomenological highlight is the significant enhancement of the diphoton decay $h \to \gamma\gamma$ at negative $\sba$, strongly correlated with one-step SFOEWPT scenarios.
Leveraging this feature, we proposed two complementary golden final states, $W^+W^-b\bar{b}\tau^+\tau^-$ and $W^+W^-b\bar{b}\gamma\gamma$, both characterized by high-multiplicity final states and negligible SM backgrounds, thus ensuring high discovery potential at the $1.5\TeV$ CLIC.

In summary, our work has revealed a distinctive phenomenology of SFOEWPTs in the inverted Type-I 2HDM. We have mapped the viable parameter space, clarified the breakdown of the vacuum uplifting correlation in multi-step transitions, highlighted the importance of multi-step dynamics for GW signatures, and identified collider channels with excellent discovery prospects. Together, these results illustrate the deep interplay between early-Universe cosmology and high-energy experiments in probing physics beyond the Standard Model.

\acknowledgments
The work of JC is supported by National Institute for Mathematical Sciences (NIMS) grant funded by the Korea government (MSIT) (No.~B25810000).
The work of  JK is supported by a KIAS Individual Grant (PG099201) at Korea Institute for Advanced Study.

\appendix

\section{Vacuum uplifting measure $\dfz$ in the Higgs basis}
\label{appendix-DF0}

For the vacuum energy different between the symmetric and broken phases at \emph{zero temperature}, it is more useful to use the Higgs basis of the 2HDM~\cite{Gunion:2002zf}, where two Higgs doublet fields $H_1$ and $H_2$ are defined by
\[
\begin{pmatrix} H_1 \\ H_2 \end{pmatrix} = 
\begin{pmatrix} \cb & \sb \\ -\sb & \cb \end{pmatrix}
\begin{pmatrix} \Phi_1 \\ \Phi_2 \end{pmatrix}.
\]
Then, $H_2$ does not develop nonzero VEV, yielding
\[
H_1 = \begin{pmatrix} G^+ \\ \frac{v + h_1 + i\,G_0}{\sqrt{2}} \end{pmatrix},
\quad
H_2 = \begin{pmatrix} H^+ \\ \frac{h_2 + i\,A}{\sqrt{2}} \end{pmatrix},
\]
where $G^+$ and $G_0$ are the Goldstone bosons to be absorbed into the SM $W^+$ and $Z$, respectively.
The tree-level potential in the Higgs basis becomes 
\begin{equation}
\label{2HDM_potential2}
\begin{split}
V_\text{tree}(H_1,H_2) &= \bar{\mu}^2_1 |H_1|^2 + \bar{\mu}^2_2 |H_2|^2 - \bar{\mu}^2 
	\bigl[ H_1^{\dagger}H_2+\hc \bigr] + \frac{\bar{\lambda}_1}{2}|H_1|^4  + \frac{\bar{\lambda}_2}{2}|H_2|^4  \\
&\quad + \bar{\lambda}_3 |H_1|^2|H_2|^2
	+ \bar{\lambda}_4 |H_1^{\dagger}H_2|^2 + \frac{\bar{\lambda}_5}{2} \bigl[ (H_1^{\dagger}H_2)^2+\hc \bigr] 
\\
&\quad + \bar{\lambda}_6 \bigl[ |H_1|^2 H_1^{\dagger}H_2+\hc \bigr]
	+ \bar{\lambda}_7 \bigl[ |H_2|^2 H_1^{\dagger}H_2+\hc \bigr].
\end{split}
\end{equation}
Let us present the parameters in \autoref{2HDM_potential2} in terms of our physical parameters $\bigl\{ \tb,\cba,M^2,v,\mh,\mhh,\ma,\mch \bigr\}$.

Using a simplified notation of $\aba$ as
\[
\aba= \sba- \cba (\tb - \tb^{-1}),
\]
the mass squared parameters are
\[
\begin{split} 
\bar{\mu}_1^2 & =  -\frac{1}{2} \bigl[ \cba^2 \mhh^2 + \sba^2 \mh^2 \bigr], \\
\bar{\mu}_2^2 & = M^2 - \frac{ \mhh^2}{2} + \frac{\mhh^2-\mh^2}{2}\sba \aba, \\
\bar{\mu}^2 & = - \frac{1}{2} (\mhh^2- \mh^2) \sba \cba.
\end{split}
\]
The modified quartic couplings are
\[
\begin{split} 
\bar{\lambda}_1 &= -\frac{ 2\bar{\mu}_1^2}{v^2}, \\
\bar{\lambda}_2 &= \frac{1}{v^2} \bigl[
	\mh^2 + (\mhh^2-M^2) (\tb-{\tb^{-1}})^2 + (m_{H_0}^2-m_{h}^2)(1-\aba^2) \bigr], \\
\bar{\lambda}_3 &= 2\frac{\mch^2-\bar{\mu}_2^2}{v^2} \\
\bar{\lambda}_4 &= \frac{1}{v^2} \bigl[
	\ma^2-2 \mch^2+ \mh^2 + (\mhh^2-\mh^2) \sba^2 \bigr], \\
\bar{\lambda}_5 &= \frac{1}{v^2} \bigl[
    -\ma^2 + \mh^2 + (\mhh^2-\mh^2) \sba^2 \bigr], \\
\bar{\lambda}_6 &= \frac{2\bar{\mu}^2}{v^2}, \\
\bar{\lambda}_7 &= \frac{1}{v^2} \bigl[
	(M^2-\mhh^2) (t_\beta-t_\beta^{-1}) - (m_{H_0}^2-m_{h}^2) \cba \aba \bigr].
\end{split}
\]

In the 2HDM, the vacuum energy difference between the 2HDM and the SM is
\begin{equation}
\label{eq-dfz}
\begin{split}	
\dfz &= \dfz^\text{tree} - \frac{\mhsm^4}{64\pi^2} (3+\log 2) 
    - \sum_{k} \frac{m_{0_k}^4}{64\pi^2} \Bigl(\log\frac{|m_{0_k}^2|}{\mu^2}-\frac{1}{2}\Bigr) \\
	&\quad + \frac{1}{64\pi^2} \sum_k \frac{1}{4} \Bigl\{
		\tilde{I}_k^2 - 2m_k^4 + \bigl[ (\tilde{I}_k-2m_k^2)^2 + m_k^2 (\tilde{J}_k-\tilde{I}_k) \bigr]
		\log\frac{m_k^2}{\mu^2} \Bigr\},
\end{split}
\end{equation}
where $k=H^\pm,A,H,h,G,G^\pm$.
The $m_{0_k}^2$ is the squared mass at the origin of the field configuration space, which are In the inverted Higgs scenario
\begin{align*}
m_{0h}^2 &= m_{0A}^2 = m_{0H^\pm}^2 = \bar{\mu_2}^2, \\
m_{0H}^2 &= m_{0G}^2 = m_{0G^\pm}^2 = \bar{\mu}_1^2.
\end{align*}
The required mass derivatives of $\tilde{I}_k$ and $\tilde{J}_k$ in \autoref{eq-dfz} are given by
\[
\begin{split}
\tilde{I}_{G} &= \mh^2 + \cba^2 (\mhh^2-\mh^2), \\
\tilde{I}_{H^\pm} &= 2\mch^2 + \cba^2 \mh^2 + \sba^2 \mhh^2 
	- \bigl[ 2M^2 - (\mhh^2-\mh^2) \mathcal{B}_{\alpha\beta} \bigr], \\
\tilde{J}_{H^\pm} &= \tilde{I}_{H^\pm} + 2 \cba^2 \sba^2 \frac{(\mhh^2-\mh^2)^2}{\mch^2}, \\
\tilde{I}_{A} &= \tilde{I}_{H^\pm} - 2\mch^2 + 2\ma^2, \\
\tilde{J}_{A} &= \tilde{I}_{A} + 2\cba^2 \sba^2 \frac{(\mhh^2-\mh^2)^2}{\ma^2}, \\
\tilde{I}_{h} &= 3\mh^2 - \cba^2 \bigl[
	2M^2 - (\mhh^2-\mh^2) \mathcal{B}_{\alpha\beta} \bigr], \\
\tilde{J}_{h} &= \tilde{I}_{h} - \frac{2\cba^2\sba^2}{(\mhh^2-\mh^2)} \bigl[
	2M^2 - (\mhh^2-\mh^2) \mathcal{B}_{\alpha\beta} \bigr]^2, \\
\tilde{I}_{H_0} &= 3\mhh^2 - \sba^2 \bigl[ 
	2M^2 - (\mhh^2-\mh^2) \mathcal{B}_{\alpha\beta} \bigr], \\
\tilde{J}_{H_0} &= \tilde{I}_{H_0} + \frac{2\cba^2\sba^2}{(\mhh^2-\mh^2)} \bigl[
	2M^2 - (\mhh^2-\mh^2) \mathcal{B}_{\alpha\beta} \bigr]^2,
\end{split}
\]
where $\mathcal{B}_{\alpha\beta} =\sa\ca/(\sb\cb)$.


\end{document}